# Cost-Effective Realization of *n*-Bit Toffoli Gates for IBM Quantum Computers Using the Bloch Sphere Approach and IBM Native Gates

ALI AL-BAYATY

Portland State University, USA, albayaty@pdx.edu, ORCID: 0000-0003-2719-0759

MAREK PERKOWSKI

Portland State University, USA, h8mp@pdx.edu, ORCID: 0000-0002-0358-1176

*Abstract –* A cost-effective *n*-bit Toffoli gate is proposed to be realized (or transpiled) based on the layouts (linear, T-like, and I-like) and the number of *n* physical qubits for IBM quantum computers. This proposed gate is termed the "layout-aware *n*-bit Toffoli gate". The layout-aware *n*-bit Toffoli gate is designed using the visual approach of the Bloch sphere, from the visual representations of the rotational quantum operations for IBM native gates. In this paper, we also proposed a new formula for the quantum cost, which calculates the total number of native gates, the crossing connections, and the depth of the final transpiled quantum circuit. This formula is termed the "transpilation quantum cost". After transpilation, our proposed layout-aware *n*-bit Toffoli gate always has a much lower transpilation quantum cost than that of the conventional *n*-bit Toffoli gate, where $3 \leq n \leq 7$ qubits, for different IBM quantum computers.

## 1 INTRODUCTION

A quantum circuit is intended to solve a (classical or quantum) problem in the quantum domain [1-7]. Such a quantum circuit is constructed using quantum gates. Generally, quantum gates are categorized into single-qubit, double-qubit, and multiple-qubit gates [8, 9]. The single-qubit gate has only one qubit to operate on. When a single-qubit gate is applied to a qubit, then the result of this operation will be reflected on the same qubit. In other words, when a quantum operation of a single-qubit gate is applied to a qubit (as its input qubit), then the result is reflected to the same qubit (as its output qubit). Single-qubit gates are categorized based on their quantum operations. These quantum operations are mainly classified into (i) the superposition, such as the Hadamard (H) gate, (ii) the fixed rotation, such as Pauli-X (X or NOT) gate, Pauli-Y (Y) gate, or Pauli-Z (Z) gate, and (iii) the customized rotation, such as the rotational X (RX) gate, the rotational Y (RY) gate, or the rotational Z (RZ) gate.

The double-qubit gate has two qubits to operate on. When a double-qubit gate is applied to one of these qubits, then the result of this operation will be reflected on the other qubit. In other words, when a quantum operation of a double-qubit gate is applied to one qubit (as its input qubit), then the result is reflected to the other qubit (as its output qubit). In quantum computing terminologies, the input qubit is termed the "control" and the output qubit is termed the "target". Such that, the quantum operation of a double-qubit gate is controlled by the control qubit, and the result is then targeted on the target qubit. Double-qubit gates are categorized based on their quantum operations. These quantum operations, in general, are classified into (i) the controlled superposition, such as the controlled-Hadamard (CH) gate, (ii) the fixed controlled rotation, such as the controlled-X (Feynman or CNOT) gate, controlled-Y (CY) gate, or controlled-Z (CZ) gate, and (iii) the customized controlled rotation, such as the controlled-RX (CRX) gate, the controlled-RY (CRY) gate, or the controlled-RZ (CRZ) gate. Note that (i) the quantum operation of a double-qubit gate is performed when the control qubit is set to the state of $|1\rangle$, and (ii) the SWAP double-qubit gate has two target qubits (without any control qubit), i.e., the SWAP gate always switches the indices of its two input qubits as swapped target qubits.

The multiple-qubit gate has $n$ qubits to operate on, where $n \geq 3$. These $n$ qubits are grouped into $n - 1$ control qubits and one target qubit. When a multiple-qubit gate is applied to the control qubits, then the result of this operation will be reflected on the target qubit. Such that, the quantum operation of a multiple-qubit gate is controlled by all control qubits. In quantum computing terminologies, the "$n$-bit" is often used instead of the "$n$ qubits" for the multiple-qubit gates. For instance, (i) a 3-bit gate has two control qubits and one target qubit, such as the controlled-CNOT (Toffoli or CCNOT) gate or the controlled-CZ (CCZ) gate, (ii) a 4-bit gate has three control qubits and one target qubit, such as the 4-bit Toffoli (CCCNOT) gate or the controlled-CCZ (CCCZ) gate, and (iii) an $n$-bit gate has $n - 1$ control qubits and one target qubit, such as the $n$-bit Toffoli gate or the $n$-bit controlled-Z gate. Note that (i) the quantum operation of a multiple-qubit gate is performed when all control qubits are set to the state of $|1\rangle$, and (ii) the controlled-SWAP (Fredkin) gate is the 3-bit gate that has one control qubit and two target qubits.

In general, a set of quantum gates (single-qubit, double-qubit, and/or multiple-qubit gates) that builds a quantum circuit is chosen based on their quantum operations to solve a (classical or quantum) problem, in the quantum domain. The quantum operations can be represented either (i) mathematically using the unitary matrices [8, 9] that numerically represent their quantum gates, or (ii) geometrically using the Bloch sphere [8, 9] that visually demonstrates the states' transitions of qubits after applying the quantum gates.

For a practical quantum application, a quantum circuit is synthesized into a quantum computer, i.e., a quantum processing unit (QPU). In IBM terminologies, the process of synthesis is termed "transpilation". Due to the technical specifications and manufacturing requirements for IBM QPUs [10-12], all multiple-qubit gates are not directly transpiled into an IBM QPU, whereas a small set of single-qubit and double-qubit gates is directly transpiled into an IBM QPU. The quantum gates of direct transpilation are termed the "native gates" [13]; otherwise, they are termed the "non-native gates". Hence, the non-native gates need to be decomposed into sets of native gates. Afterward, all native gates, sets of native gates, and their qubits are mapped into the layout (or topology) of an IBM QPU. Note that (i) the decomposition and the mapping steps are parts of the transpilation process, and (ii) a layout is the architecture (or the geometrical placement) of the physical qubits for an IBM QPU. For the IBM quantum system, the native single-qubit gates are I, X, $\sqrt{X}$ (RX of $\pi/2$ radians or V [14]), and RZ, as well as there is only one native double-qubit gate, which is the CNOT (Feynman) gate.

In this paper, we utilized three IBM QPUs of different layouts and the number of physical qubits, to perform our research in transpiling the non-native $n$-bit Toffoli gate, which is also called the "conventional $n$-bit Toffoli gate", where $3 \leq n \leq 7$ qubits. The utilized IBM QPUs are: (i) the `ibmq_manila` of linear layout for five qubits [15], (ii) the `ibmq_quito` of T-like layout for five qubits [16], and (iii) the `ibmq_perth` of I-like layout for seven qubits [10].

In our research, we noticed the following complications after transpiling the conventional $n$-bit Toffoli gate, using the aforementioned IBM QPUs and the IBM-based transpilation software:

1. A long-decomposed set of native (single-qubit and double-qubit) gates is generated.
2. Too many crossing connections are added, i.e., SWAP gates are added among the physical qubits, and the complexity of mapping into the layout of an IBM QPU is raised.
3. The depth (as the critical longest path through the total number of native gates) is increased, which produces a longer delay for the transpiled quantum circuit and increases the decoherence for the physical qubits [10-12].

From these complications, the quantum cost of the final transpiled quantum circuit is dramatically increased as $n$ increases! In our research, for ease of calculations and analyses, we introduced the term "transpilation quantum cost", which calculates the total number of (i) the native single-qubit gates, (ii) the native double-qubit gates, (iii) the crossing connections, and (iv) the depth of the final transpiled quantum circuit.

In this paper, we proposed a cost-effective $n$-bit Toffoli gate that can be properly transpiled into different layouts of IBM QPUs. This cost-effective $n$-bit Toffoli gate has a lower transpilation quantum cost than that of the conventional $n$-bit Toffoli gate after transpilation. This cost-effective $n$-bit Toffoli gate is termed the "layout-aware $n$-bit Toffoli gate". The layout-aware $n$-bit Toffoli gate consists of:

1. Fewer native (single-qubit and double-qubit) gates are generated, for $3 \leq n \leq 7$.
2. No crossing connections, for $3 \leq n \leq 4$, i.e., no SWAP gates are added among the physical qubits, and a simple structure is generated to decrease the complexity of mapping.
3. Reduced depth, for $3 \leq n \leq 7$, i.e., shorter delay and the decoherence is decreased.

The layout-aware $n$-bit Toffoli gate is designed to be utilized as a transpilation software package for the IBM quantum system. For instance, when a researcher transpiles a quantum circuit consisting of many conventional $n$-bit Toffoli gates, this transpilation software package will transpile these many conventional gates to many layout-aware $n$-bit Toffoli gates, for a lower transpilation quantum cost based on the layout of an IBM QPU.

In our research, various experiments of conventional $n$-bit Toffoli gates and layout-aware $n$-bit Toffoli gates are designed and transpiled, for $3 \leq n \leq 7$. After transpiling these experiments using the previously stated IBM QPUs, we observed that all layout-aware $n$-bit Toffoli gates (using our software package) have lower transpilation quantum costs than that of the conventional $n$-bit Toffoli gates (using the IBM-based transpilation software). Note that the utilized IBM-based transpilation software is the "Transpiler" package [17].

Different designs and implementations were proposed to decompose the conventional $n$-bit Toffoli gate into a set of single-qubit and double-qubit gates and/or a set of conventional 3-bit Toffoli gates. Barenco *et al.* [14] investigated different approaches to decompose the conventional 3-bit Toffoli gate into a set of single-qubit gates (RY) and double-qubit gates (controlled-V and controlled-V$^\dagger$), also the authors decomposed the conventional $n$-bit Toffoli gate into a set of conventional 3-bit Toffoli gates. Shende and Markov [18] decomposed the conventional 3-bit Toffoli gate into several single-qubit gates (H, T, and T$^\dagger$) with less than six CNOT gates, as well as they decomposed the CNOT gates into the CZ gates. Schmitt and De Micheli [19] presented the "Tweedledum" as the open-source library that aims to link high-level algorithms and QPUs, also the authors demonstrated the decomposition of conventional 3-bit Toffoli gate using the approaches of Decision Diagrams and Phase polynomials through the use of single-qubit gates (T and T$^\dagger$) and double-qubit gates (CNOT). Tan and Cong [20] discussed different layouts synthesizing for transforming quantum circuits to meet quantum hardware limitations, the authors' two synthesizers (optimal and approximate) are introduced to reduce the depth and the number of SWAP gates for different QPUs (IBM QX2, IBM Melbourne, and Rigetti Aspen-4). Lukac *et al.* [21] proposed a method that maps a quantum circuit to the layout of a QPU, by using the circuit interaction graph (CIG) that is reduced to minimize the number of SWAP gates and the path length between the neighboring physical qubits of different layouts of IBM QPUs (Tenerife, Melbourne, Ruschlikon, and Tokyo). From these papers, we noticed that all decompositions of the conventional $n$-bit Toffoli gates have high transpilation quantum costs for different layouts of IBM QPUs, since (i) the non-native gates need to be further decomposed, (ii) many SWAP gates are added, and (iii) the final depth is dramatically increased as $n$ increases. In this paper, our experiments prove that the proposed layout-aware $n$-bit Toffoli gate always has a much lower transpilation quantum cost for different layouts of IBM QPUs, for $3 \leq n \leq 7$ qubits.

## 2 DESIGN AND ANALYSIS

The Bloch sphere is the three-dimensional geometrical sphere of three axes (X, Y, and Z) that represents the quantum state of a qubit. When a quantum gate is applied to a qubit, the Bloch sphere visualizes this quantum operation in Hilbert space ($\mathfrak{H}$) [8, 9]. As depicted in Figure 1, the quantum state of a qubit is visualized on the three axes of the Bloch sphere (note that no quantum operation has been applied yet):

- The Z-axis represents the pure state, or the computational basis state, of a qubit for the base vector of $|0\rangle$ or $|1\rangle$ using Dirac notation [8, 9]; also denoted by $\begin{bmatrix}1\\0\end{bmatrix}$ or $\begin{bmatrix}0\\1\end{bmatrix}$ using Heisenberg notation [8, 9], respectively. The Z-axis is also used for the final measurement of a quantum circuit, in the classical domain.
- The X-axis represents the pure superposition state of a qubit for $|+\rangle = \frac{1}{\sqrt{2}}(|0\rangle + |1\rangle)$ or $|-\rangle = \frac{1}{\sqrt{2}}(|0\rangle - |1\rangle)$; also denoted by $\frac{1}{\sqrt{2}}\begin{bmatrix}1\\1\end{bmatrix}$ or $\frac{1}{\sqrt{2}}\begin{bmatrix}1\\-1\end{bmatrix}$, respectively.
- The Y-axis represents the rotational superposition state of a qubit for $|+i\rangle = \frac{1}{\sqrt{2}}(|0\rangle + i|1\rangle)$ or $|-i\rangle = \frac{1}{\sqrt{2}}(|0\rangle - i|1\rangle)$; also denoted by $\frac{1}{\sqrt{2}}\begin{bmatrix}1\\i\end{bmatrix}$ or $\frac{1}{\sqrt{2}}\begin{bmatrix}1\\-i\end{bmatrix}$, respectively.

In this paper, Dirac notation will be employed for the Bloch sphere visualization, the state's transitions of qubits, and the analysis of quantum circuits. For ease of illustrating the unitary matrices (numerical representations) [8, 9], Heisenberg notation will be utilized for building the mathematical models of our proposed layout-aware *n*-bit Toffoli gates.

### 2.1 The XY-Plane

A quantum gate (single-qubit, double-qubit, or multiple-qubit), which has X, Y, or Z in its notation, rotates the state of a qubit around the X-, Y-, or Z-axis of the Bloch sphere, respectively. Such a rotation around an axis of the Bloch sphere is dependent on a defined rotational angle ($\theta$). Figure 1 demonstrates the Bloch sphere and its three axes with their respective rotational angles, where the ($+\theta$) represents the counterclockwise rotation and the ($-\theta$) represents the clockwise rotation. In this paper, for the same angular degree, the ($\theta$) represents the counterclockwise rotation and the ($\Phi$) represents the clockwise rotation, as stated in (1) below.

$$\Phi = -\theta \qquad (1)$$

For instance,

- The Z gate rotates the state of a qubit around the Z-axis by a $\theta$ of $\pi$ radians, such that the Z gate is equivalent to the RZ($\pi$) gate.
- The $\sqrt[2]{Z}$ gate rotates the state of a qubit around the Z-axis by a $\theta$ of $\pi/2$ radians, such that the $\sqrt[2]{Z}$ gate is equivalent to the RZ($\pi/2$) gate. Note that the $\sqrt[2]{Z}$ gate is also termed the S gate [8, 9, 14].
- The $\sqrt[4]{Z}$ gate rotates the state of a qubit around the Z-axis by a $\theta$ of $\pi/4$ radians, such that the $\sqrt[4]{Z}$ gate is equivalent to the RZ($\pi/4$) gate. Note that the $\sqrt[4]{Z}$ gate is also termed the T gate [8, 9, 19].

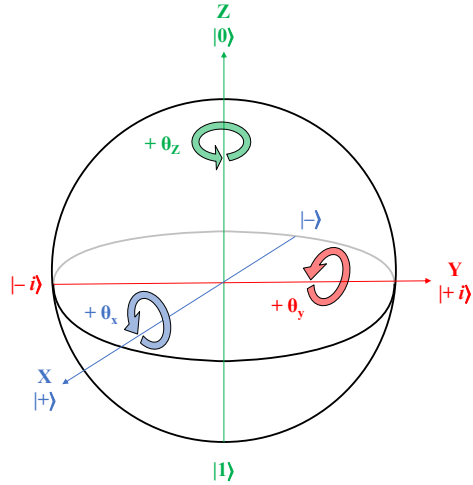

Figure 1: Schematic of the Bloch sphere consists of (i) three axes (X in blue, Y in red, and Z in green), (ii) the quantum state of a qubit on these axes individually, and (iii) the respective rotational angles for these three axes ($\theta_x$, $\theta_y$, and $\theta_z$). Note that the (+ $\theta$) represents the counterclockwise angular rotation and the (– $\theta$) represents the clockwise angular rotation for their respective axes.

All quantum gates (single-qubit, double-qubit, and multiple-qubit) are unitary gates [8, 9], and a few of them are non-Hermitian gates [8, 9], i.e., their quantum operations are not in their own inverses. For instance, the $S^{\dagger}$ or RZ(– $\pi$/2) is the inverse gate for the S gate, and the $T^{\dagger}$ or RZ(– $\pi$/4) is the inverse gate for the T gate. Note that some rotational gates alter the state's phase of a qubit, such that the choice of rotational gates is a critical factor in the design of a quantum circuit.

In general, the Toffoli gate is the quantum counterpart of the classical Boolean AND gate. Therefore, when all controls (input qubits) are set to the state of $|1\rangle$, then the Toffoli gate flips the target (output qubit) to the state of $|1\rangle$ that was initially in the state of $|0\rangle$; otherwise, the target remains at its initial state of $|0\rangle$.

For IBM QPUs [10, 14, 15], the RZ gate is the native single-qubit gate, and its quantum operation (as rotation) is performed on the XY-plane of the Bloch sphere, i.e., around the Z-axis. When a qubit was initially set to the state of $|+\rangle$, $|-\rangle$, $|+i\rangle$, or $|-i\rangle$, Figure 2 illustrates the XY-plane (as the top-view of the Bloch sphere) and its placements (circles and squares), where:

- The circles represent the placements after applying the RZ(+ $\pi$/2) or RZ(– $\pi$/2) gate to a qubit.
- The squares represent the placements after applying the RZ(+ $\pi$/4) or RZ(– $\pi$/4) gate to a qubit.

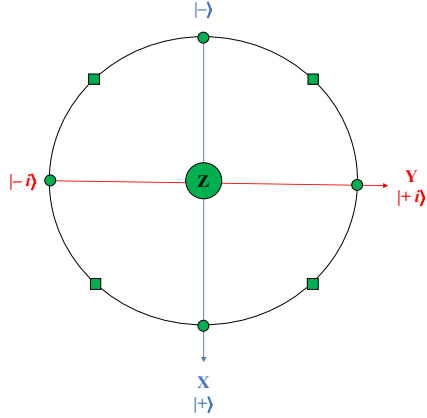

Figure 2: Schematic of the XY-plane (as the top-view) of the Bloch sphere, where (i) the circles represent the placements of the state of a qubit after applying RZ(+ $\pi/2$) or RZ(− $\pi/2$) gate, and (ii) the squares represent the placements of the state of a qubit after applying RZ(+ $\pi/4$) or RZ(− $\pi/4$) gate. Note that such a qubit was initially set to the state of $|+\rangle$, $|-\rangle$, $|+i\rangle$, or $|-i\rangle$.

## 2.2 Geometrical Design Approach

The layout-aware *n*-bit Toffoli gate geometrically operates on the XY-plane of the Bloch sphere, and it is only designed by utilizing IBM native gates, to be cost-effectively transpiled into IBM QPUs. The utilized native gates are $\sqrt{X}$, RZ, and CNOT gates, where:

- $\sqrt{X}$ and RZ gates compose the non-native Hadamard (H) gates, which place the target (output qubit) of the layout-aware *n*-bit Toffoli gate to/from the XY-plane.
- CNOT gates rotate the state of the target qubit around the X-axis of the XY-plane, and CNOT gates are controlled by the states of controls (input qubits) of the layout-aware *n*-bit Toffoli gate.
- RZ gates, additionally, rotate the state of the target qubit in the XY-plane, based on the decisions of CNOT gates, i.e., based on the states of controls.

Note that, from this geometrical design approach: (i) the target qubit is always initially set to the state of $|0\rangle$, and (ii) the mostly utilized native gates are the single-qubit RZ gates. The cost-effective transpilation is guaranteed, due to: (i) the decomposition step is not required, and (ii) the number of native single-qubit gates is larger than that of the native double-qubit gates.

Moreover, two H gates are required to place the target qubit to/from the XY-plane of the Bloch sphere. The first H gate transforms the target qubit of the initial state of $|0\rangle$ to the state of $|+\rangle$, i.e., moving the target from the Z-axis to the X-axis of the XY-plane. The state of the target qubit rotates from $|+\rangle$ to $|-\rangle$ when all controls are in the state of $|1\rangle$; otherwise, the state of the target qubit remains at $|+\rangle$ state when any of the controls is in the state of $|0\rangle$. The second H gate transforms the final state of the target qubit from the XY-plane back to the Z-axis (either to $|0\rangle$ or to $|1\rangle$) depending on the states of the control qubits.

In such a scenario, the two H gates will surround the quantum rotational operations of the layout-aware *n*-bit Toffoli gate. In this paper, the quantum rotational operations of layout-aware *n*-bit Toffoli gate (that are surrounded by H gates) are termed the "core", as stated in (2) below.

$$\text{Layout-aware } n\text{-bit Toffoli gate} = H \ \{ \ \text{Core} \ \} \ H \qquad (2)$$

***Definition 1.*** *The core of the layout-aware n-bit Toffoli gate is a set of IBM native (RZ and CNOT) gates, when the two Hadamard (H) gates are excluded, where the H gate is composed of IBM native ($\sqrt{X}$ and RZ) gates, and $n \geq 3$ qubits* □

However, the non-native H gate is composed of three native gates (one $\sqrt{X}$ and two RZ), as expressed in (3) below. Hence, the layout-aware *n*-bit Toffoli gate is entirely constructed from IBM's native gates of $\sqrt{X}$, RZ, and CNOT, as stated in (4) below, and it can be cost-effectively transpiled to an IBM QPU. Note that, in (4), the second decomposed H gate is the inverse quantum rotational operation as compared to the first decomposed H gate. Therefore, the layout-aware *n*-bit Toffoli gate is unitary.

$$H = RZ\left(\frac{\pi}{2}\right) \sqrt{X} \; RZ\left(\frac{\pi}{2}\right) \tag{3}$$

$$\text{Layout-aware } n\text{-bit Toffoli gate} = \underbrace{RZ\left(\frac{\pi}{2}\right) \sqrt{X} \; RZ\left(\frac{\pi}{2}\right)}_{\text{1}^{st} \text{ decomposed H gate}} \{\text{Core}\} \underbrace{RZ\left(\frac{\pi}{2}\right) \sqrt{X} \; RZ\left(\frac{\pi}{2}\right)}_{\text{2}^{nd} \text{ decomposed H gate}} \tag{4}$$

In our work, the layout-aware *n*-bit Toffoli gate is designed as a quantum circuit of straight mirrored hierarchical structure, such that:

- The straight structure eliminates the crossing connections among the qubits; thus, no SWAP gates are added and the complexity of mapping the final transpiled quantum circuit into an IBM QPU is decreased.
- The mirrored structure ensures that the layout-aware *n*-bit Toffoli gate is constructed symmetrically; thus, when any of the controls are in the state of |0⟩, i.e., there is a false value of the solution, then all quantum circuit is collapsed to generate a state of |0⟩ for the target qubit.
- The hierarchical structure maintains the generation of layout-aware *n*-bit Toffoli gate from the repetition of layout-aware *x*-bit Toffoli gates, where $3 \leq x \leq n-1$; thus, the layout-aware *n*-bit Toffoli gate is hierarchically generated from the layout-aware ($n-1$)-bit Toffoli gate, then the layout-aware ($n-2$)-bit Toffoli gate, and so on until the layout-aware 3-bit Toffoli gate.

Note that (i) the symmetrical construction means that a quantum circuit has identical numbers and types of quantum gates on its left-side and right-side, and (ii) the collapsing of a quantum circuit means that the quantum gates on either side will cancel the symmetrical quantum gates on the other side [8, 9, 22].

***Definition 2.*** *The layout-aware n-bit Toffoli gate is a quantum circuit of straight mirrored hierarchical structure for m controls (input qubits) and one target (output qubit), where $m = n - 1$ and $n \geq 3$ qubits. This quantum circuit is only constructed from IBM native single-qubit gates ($\sqrt{X}$ and RZ) and native double-qubit gates (CNOT), to ensure the cost-effective transpilation into an IBM QPU* □

The core of the layout-aware 3-bit Toffoli gate is the main quantum circuit that is utilized to iteratively build the core of the layout-aware *n*-bit Toffoli gate. For instance, the core of the layout-aware 3-bit Toffoli gate is mirrored and repeated to construct the core of the layout-aware 4-bit Toffoli gate. Subsequently, the core of the layout-aware 4-bit Toffoli gate is mirrored and repeated to construct the core of the layout-aware 5-bit Toffoli gate, and so on for the core of the layout-aware *n*-bit Toffoli gate. Finally, the first and second H gates are added to the core of the layout-aware *n*-bit Toffoli gate to generate the final quantum circuit of the layout-aware *n*-bit Toffoli gate.

For the layout-aware *n*-bit Toffoli gate, the false solution (as the target qubit is in the state of |0⟩) occurs when any of the controls are in the state of |0⟩; otherwise, the true solution (as the target qubit is in the state of |1⟩) occurs when all controls are in the state of |1⟩.

For instance, for the layout-aware *n*-bit Toffoli gate of a false solution (the target qubit remains in the |0⟩ state):

1. The mirrored and repetitive hierarchical core of layout-aware *n*-bit Toffoli gate collapses to the core of layout-aware (*n*–1)-bit Toffoli gate.
2. The mirrored and repetitive hierarchical core of layout-aware (*n*–1)-bit Toffoli gate collapses to the core of layout-aware (*n*–2)-bit Toffoli gate, and so on until *n* = 3.
3. The mirrored and repetitive hierarchical core of the layout-aware 3-bit Toffoli gate collapses to cancel itself.
4. Finally, the first and second H gates cancel each other to generate the state of |0⟩, as the false solution, for the target qubit that is initially set to the state of |0⟩.

For the core of layout-aware 3-bit Toffoli gate, the rotational angles ($\theta$ and $\Phi$) of RZ gates are accurately chosen, where $\theta = \pi/4$ radians and $\Phi = -\pi/4$ radians. Such that, when a false solution occurs, RZ gates will cancel each other. On the other hand, when a true solution occurs, the $\pm \pi/4$ radians are chosen to cover the lower and upper halves of the XY-plane of the Bloch sphere. The complete construction of the layout-aware 3-bit Toffoli gate is expressed in (5) below. Figure 3 demonstrates the quantum circuit of the layout-aware 3-bit Toffoli gate, and Figure 4 geometrically illustrates the angular rotations of the target qubit in the XY-plane when both controls are in the state of |1⟩, i.e., a true solution occurs. For ease of illustration and notation, in this subsection and the following ones, the H gates are not decomposed into their IBM native gates, to save space.

$$\text{Layout-aware 3-bit Toffoli gate} = H \underbrace{\{(RZ_\theta \, X_{C_1} RZ_\Phi) \, X_{C_0} \, (RZ_\theta \, X_{C_1} RZ_\Phi)\}}_{\text{Core of layout-aware 3-bit Toffoli gate}} H \qquad (5)$$

Where,
$\theta = \pi/4$ radians $= -\Phi$ for RZ gates ($RZ_\theta$ and $RZ_\Phi$),
$C_0$ and $C_1$ are the controls (input qubits),
$X_{C_0}$ is the CNOT gate that exists when $C_0 = |1⟩$, and
$X_{C_1}$ is the CNOT gate that exists when $C_1 = |1⟩$.

As depicted in Figure 3, due to the straight structure of the layout-aware 3-bit Toffoli gate, i.e., there are no crossing connections among all qubits, the target qubit can be switched with any of the control qubits for cost-effective mapping to any layout of an IBM QPU.

From Figure 3 and Figure 4, for all states of control qubits, Table 1 stated the phase transitions logic of the states of the target qubit, for the layout-aware 3-bit Toffoli gate, where N.A. means "not applicable".

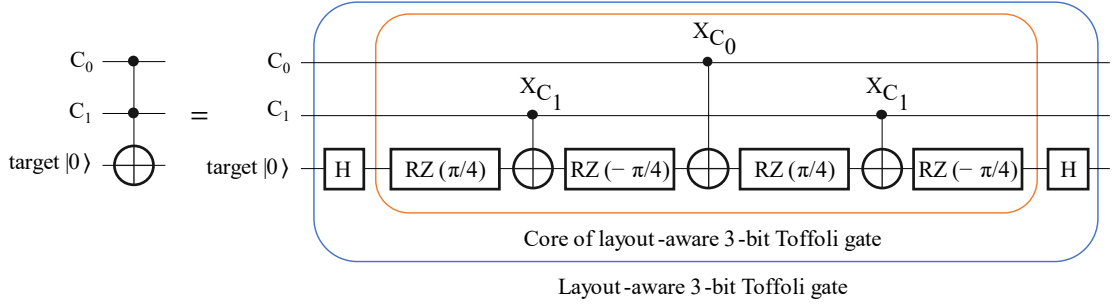

Figure 3: Schematic of the quantum circuits of (i) the conventional 3-bit Toffoli gate (left-side), (ii) the core of layout-aware 3-bit Toffoli gate (in the orange area – right-side), and (iii) the layout-aware 3-bit Toffoli gate (in the blue area – right-side), where the rotational angles $\theta = \pi/4$ radians $= -\Phi$ for RZ gates, and the target qubit of layout-aware 3-bit Toffoli gate can be switched with any of the control qubits to be mapped to any layout of an IBM QPU.

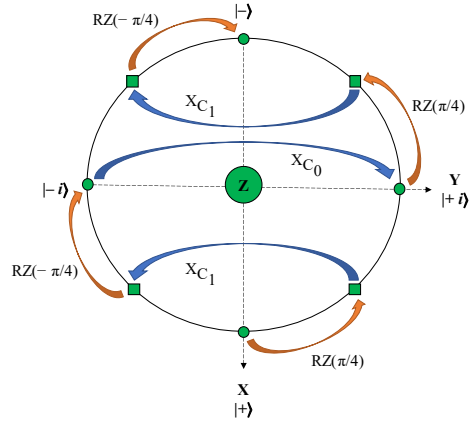

Figure 4: Geometrical representation of the XY-plane of the Bloch sphere, for the rotations of the target qubit for the core of layout-aware 3-bit Toffoli gate, where (i) the target qubit is in the state of $|+\rangle$ after applying the first H gate, (ii) both controls ($C_0$ and $C_1$) are in the state of $|1\rangle$, (iii) CNOT gates ($X_{C_0}$ and $X_{C_1}$) are represented by the blue arrows, (iv) $RZ_\theta$ and $RZ_\Phi$ gates are represented by the orange arrows, and (iv) $\theta = \pi/4$ radians $= -\Phi$. When the target qubit is in the state of $|-\rangle$, the second H gate transforms it to the state of $|1\rangle$, to represent the true solution.

Table 1: Phase transitions logic of the states of target qubit for all states of control qubits, for the layout-aware 3-bit Toffoli gate.

| Controls | Phases and states of the target (initially set to $\|0\rangle$) after applying the gates | | | | | | | | | Solutions in Boolean logic values |
|---|---|---|---|---|---|---|---|---|---|---|
| $\|C_1 C_0\rangle$ | H | RZ ($\pi/4$) | $X_{C_1}$ | RZ ($-\pi/4$) | $X_{C_0}$ | RZ ($\pi/4$) | $X_{C_1}$ | RZ ($-\pi/4$) | H | |
| $\|0\,0\rangle$ | $\|+\rangle$ | $\pi/4\,\|+\rangle$ | N.A. | $\|+\rangle$ | N.A. | $\pi/4\,\|+\rangle$ | N.A. | $\|+\rangle$ | $\|0\rangle$ | False |
| $\|0\,1\rangle$ | $\|+\rangle$ | $\pi/4\,\|+\rangle$ | N.A. | $\|+\rangle$ | $\|+\rangle$ | $\pi/4\,\|+\rangle$ | N.A. | $\|+\rangle$ | $\|0\rangle$ | False |
| $\|1\,0\rangle$ | $\|+\rangle$ | $\pi/4\,\|+\rangle$ | $-\pi/4\,\|+\rangle$ | $\|-i\rangle$ | N.A. | $-\pi/4\,\|+\rangle$ | $\pi/4\,\|+\rangle$ | $\|+\rangle$ | $\|0\rangle$ | False |
| $\|1\,1\rangle$ | $\|+\rangle$ | $\pi/4\,\|+\rangle$ | $-\pi/4\,\|+\rangle$ | $\|-i\rangle$ | $\|+i\rangle$ | $3\pi/4\,\|+\rangle$ | $-3\pi/4\,\|+\rangle$ | $\|-\rangle$ | $\|1\rangle$ | True |

Table 1 expresses the conclusion that the layout-aware 3-bit Toffoli gate acts as the quantum counterpart of the classical Boolean 2-input AND gate.

As previously shown in Figure 3, there are no crossing connections among the qubits in the design of the layout-aware 3-bit Toffoli gate. Due to the mechanisms of straight mirrored hierarchical structure:

- There are not any crossing connections for the layout-aware 4-bit Toffoli gate, and so on until the layout-aware $n$-bit Toffoli gate.
- From (5) above, the layout-aware 3-bit Toffoli gate can also be expressed by reversing the order of rotational angles ($\theta$ and $\Phi$) for RZ gates, as presented in (6) below. Note that the rotational operations on the XY-plane are revered and the phase transition logic is flipped as well.

$$\text{Layout-aware 3-bit Toffoli gate} = H \; \underbrace{\{(RZ_\Phi \; X_{C_1} RZ_\theta) \; X_{C_0} \; (RZ_\Phi \; X_{C_1} RZ_\theta)\}}_{\text{Core of layout-aware 3-bit Toffoli gate}} \; H \qquad (6)$$

When the core of the layout-aware 3-bit Toffoli gate is mirrored and an additional CNOT gate is inserted between these cores, and this additional CNOT gate is controlled by an additional control, then the core of the layout-aware 4-bit Toffoli gate is constructed, as illustrated in Figure 5.

For the core of layout-aware 4-bit Toffoli gate, to rotate the target qubit in the XY-plane of the Bloch sphere for a false solution $|+\rangle$ or a true solution $|-\rangle$, the rotational angles ($\theta$ and $\Phi$) of RZ gates need to be reduced to ($\pi/8$) and ($-\pi/8$) radians, respectively. In other words, the core of the layout-aware 4-bit Toffoli gate consists of two cores of layout-aware 3-bit Toffoli gate and one additional control, such that:

- The first core of the layout-aware 3-bit Toffoli gate occupies the lower half of the XY-plane, and the rotational angles are reduced by 50% from ($\theta = \pi/4 = -\Phi$) to ($\theta = \pi/8 = -\Phi$).
- The second core of the layout-aware 3-bit Toffoli gate occupies the upper half of the XY-plane, and the rotational angles are reduced by 50% from ($\theta = \pi/4 = -\Phi$) to ($\theta = \pi/8 = -\Phi$).
- The additional control (as the CNOT gate) switches between the upper and the lower halves.

For ease of design and illustration, the controls ($C_0$ and $C_1$) are shifted to the controls ($C_1$ and $C_2$), respectively, whereas the additional control becomes $C_0$, as stated in (7) below and demonstrated in Figure 5. The layout-aware 4-bit Toffoli gate with the respective geometrical design approach for the XY-plane of the Bloch sphere is illustrated in Figure 6.

$$\text{Layout-aware 4-bit Toffoli gate} = H \; \underbrace{\left\{ \begin{pmatrix} \text{Core of layout-aware} \\ \text{3-bit Toffoli gate} \\ (\theta = \pi/8 = -\Phi) \end{pmatrix} X_{C_0} \begin{pmatrix} \text{Core of layout-aware} \\ \text{3-bit Toffoli gate} \\ (\theta = \pi/8 = -\Phi) \end{pmatrix} \right\}}_{\text{Core of layout-aware 4-bit Toffoli gate}} \; H \qquad (7)$$

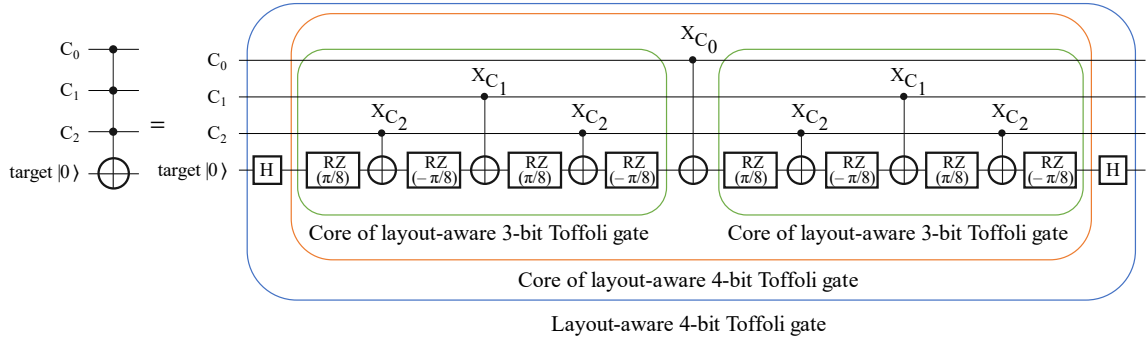

Layout-aware 4-bit Toffoli gate

Figure 5: Schematic of the quantum circuits of (i) the conventional 4-bit Toffoli gate (left-side), (ii) the two cores of layout-aware 3-bit Toffoli gate (in the green areas – right-side), (iii) the core of layout-aware 4-bit Toffoli gate (in the orange area – right-side), and (iv) the layout-aware 4-bit Toffoli gate (in the blue area – right-side), where the rotational angles θ = π/8 radians = – Φ for RZ gates, and the target qubit of layout-aware 4-bit Toffoli gate can be switched with any of the control qubits to be mapped to any layout of an IBM QPU.

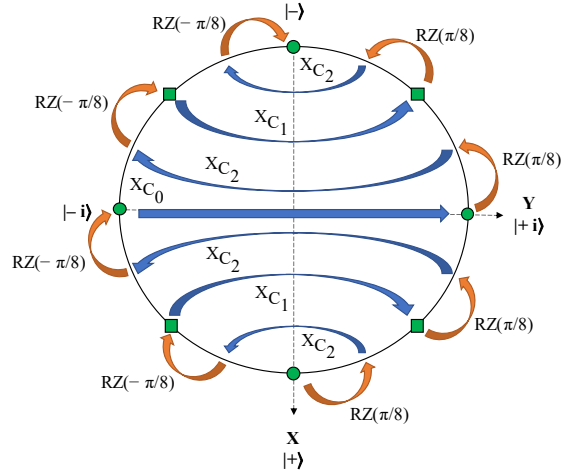

Figure 6: Geometrical representation of the XY-plane of the Bloch sphere, for the rotations of the target qubit for the core of layout-aware 4-bit Toffoli gate, where (i) the target qubit is in the state of |+⟩ after applying the first H gate, (ii) All controls ($C_0$, $C_1$, and $C_2$) are in the state of |1⟩, (iii) CNOT gates ($X_{C_0}$, $X_{C_1}$, and $X_{C_2}$) are represented by the blue arrows, (iv) $RZ_θ$ and $RZ_Φ$ gates are represented by the orange arrows, (iv) θ = π/8 radians = – Φ. When the target qubit is in the state of |–⟩, the second H gate transforms it to the state of |1⟩, to represent the true solution.

In a similar scenario, the core of the layout-aware 4-bit Toffoli gate is mirrored with an additional CNOT gate inserted between these cores, and the core of the layout-aware 5-bit Toffoli is constructed. The controls ($C_0$, $C_1$, and $C_2$) of the core of the layout-aware 4-bit Toffoli gate are shifted to the controls ($C_1$, $C_2$, and $C_3$), respectively, and the additional control becomes $C_0$. Moreover, the rotational angles (θ and Φ) of RZ gates are reduced to (π/16) and (– π/16), respectively, to rotate the target qubit in the lower and the upper halves of the XY-plane of the Bloch sphere.

Subsequently, the core of the layout-aware $n$-bit Toffoli gate is constructed by mirroring the core of the layout-aware ($n$–1)-bit Toffoli gate and inserting the additional CNOT gate, as expressed in (8) below. To rotate the target qubit in the lower and the upper halves of the XY-plane, the new rotational angles ($\theta$ and $\Phi$) of RZ gates are reduced based on $m$ qubits as stated in (9) below, where $m$ is the number of control qubits, $m = n - 1$, and $n \geq 3$. Figure 7 demonstrates the quantum circuit of the layout-aware $n$-bit Toffoli gate. Note that (i) the final construction of the layout-aware $n$-bit Toffoli gate consists of $2^{(n-3)}$ cores of layout-aware 3-bit Toffoli gate plus two H gates, and (ii) all cores of the layout-aware 3-bit Toffoli gate until the layout-aware $n$-bit Toffoli gate are generated as the expanded structure of balanced binary tree [23].

$$\text{Layout-aware } n\text{-bit Toffoli gate} = H \underbrace{\left\{ \begin{Bmatrix} \text{Core of layout-aware} \\ (n\text{–1})\text{-bit Toffoli gate} \\ (\theta = -\Phi) \end{Bmatrix} X_{C_0} \begin{Bmatrix} \text{Core of layout-aware} \\ (n\text{–1})\text{-bit Toffoli gate} \\ (\theta = -\Phi) \end{Bmatrix} \right\}}_{\text{Core of layout-aware } n\text{-bit Toffoli gate}} H \qquad (8)$$

$$\theta = \pi / 2^m = -\Phi \qquad (9)$$

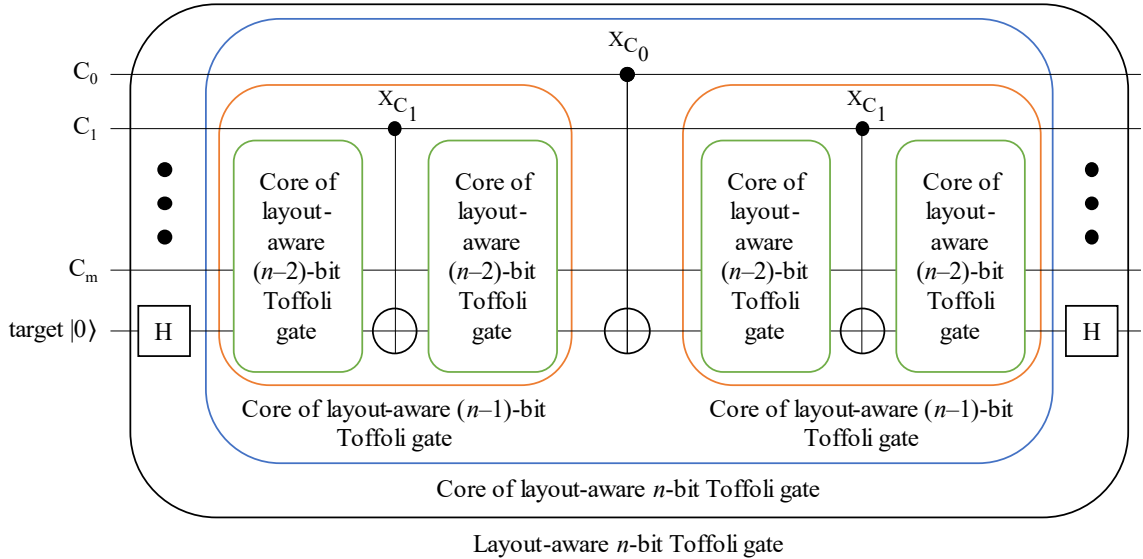

Figure 7: Schematic of the quantum circuit of (i) four cores of layout-aware ($n$–2)-bit Toffoli gate (in green areas), (ii) two cores of layout-aware ($n$–1)-bit Toffoli gate (in orange areas), (iii) the core of layout-aware $n$-bit Toffoli gate (in the blue area), (iv) layout-aware $n$-bit Toffoli gate (in the black area), and (v) the layout-aware $n$-bit Toffoli gate consists of $2^{(n-3)}$ cores of layout-aware 3-bit Toffoli gate plus two H gates, where the rotational angles $\theta = \pi/2^m$ radians $= -\Phi$ for RZ gates, $m$ is the number of control qubits, $m = n - 1$, $n \geq 3$, and the target qubit of layout-aware $n$-bit Toffoli gate can be switched with any of the control qubits to be mapped to any layout of an IBM QPU.

From (8) and (9) above, and as illustrated in Figure 7, the construction of the layout-aware $n$-bit Toffoli gate can be expressed algorithmically as stated in ALGORITHM 1. The symbol ($\rightarrow$) means "apply to". This algorithm constructs the completed quantum circuit of layout-aware $n$-bit Toffoli gate as the balanced binary tree.

| | ALGORITHM 1: Layout-aware *n*-bit Toffoli gate | |
|---|---|---|
| | **Inputs:** list of *m* indexed controls (input qubits) and one indexed target (output qubit) | |
| | **Output:** the quantum circuit of layout-aware *n*-bit Toffoli gate | |
| 1. | $n = m + 1$ | ► Total number of qubits (*m* controls + 1 target) |
| 2. | $\theta = \pi / 2^m$ | ► Counterclockwise rotational angle |
| 3. | $\Phi = -\theta$ | ► Clockwise rotational angle |
| 4. | **Construct** Core of layout-aware 3-bit Toffoli gate **of:** | |
| 5. | One indexed target | |
| 6. | Two controls of indices $[m-1, m]$ | |
| 7. | Three CNOT gates | ► From the indexed controls to the indexed target |
| 8. | Four RZ gates | ► Of rotational angles ($\theta$ and $\Phi$) for the indexed target |
| 9. | **End Construct** | |
| 10. | $r = m - 2$ | ► Calculate the remaining indexed controls |
| 11. | **While** ($r > 0$) **do:** | |
| 12. | *backup* = the previously constructed core | |
| 13. | **Insert** CNOT gate | ► From the indexed control [*r*] to the indexed target |
| 14. | **Append** *backup* | ► The mirrored structure of the core |
| 15. | $r = r - 1$ | ► The next remaining indexed control |
| 16. | **End While** | |
| 17. | H → *target* | ► The first Hadamard gate |
| 18. | H → *target* | ► The second Hadamard gate |

### 2.3 Mathematical Models

From (5) above, the layout-aware 3-bit Toffoli gate, of two control qubits ($C_0$ and $C_1$) and one target qubit (t), is constructed from a set of IBM native gates:

- Four native single-qubit gates (RZ).
- Three native double-qubit (CNOT).
- Two H gates (four RZ and two $\sqrt{X}$).

By using the unitary matrices (numerical representations) of these quantum gates, the layout-aware 3-bit Toffoli gate is mathematically modeled in (10) below. For ease of notation, the unitary matrices for all gates are expressed in the dimensions of 2×2, instead of using Kronecker ($\otimes$) notation for matrices multiplication [8, 9].

Layout-aware 3-bit Toffoli gate $= \frac{1}{\sqrt{2}}\begin{bmatrix}1 & 1\\1 & -1\end{bmatrix} \left\{ \left(\begin{bmatrix}1 & 0\\0 & e^{i\theta}\end{bmatrix}\begin{bmatrix}0 & 1\\1 & 0\end{bmatrix}_{C_1}\begin{bmatrix}1 & 0\\0 & e^{i\Phi}\end{bmatrix}\right)\begin{bmatrix}0 & 1\\1 & 0\end{bmatrix}_{C_0}\left(\begin{bmatrix}1 & 0\\0 & e^{i\theta}\end{bmatrix}\begin{bmatrix}0 & 1\\1 & 0\end{bmatrix}_{C_1}\begin{bmatrix}1 & 0\\0 & e^{i\Phi}\end{bmatrix}\right)\right\}$ $\frac{1}{\sqrt{2}}\begin{bmatrix}1 & 1\\1 & -1\end{bmatrix}$

$= \frac{1}{2}\left(\begin{bmatrix}1 & 1\\1 & -1\end{bmatrix}\left\{\left(\begin{bmatrix}0 & e^{i\theta}\\e^{i\Phi} & 0\end{bmatrix}_{C_1}\right)\begin{bmatrix}0 & 1\\1 & 0\end{bmatrix}_{C_0}\left(\begin{bmatrix}0 & e^{i\theta}\\e^{i\Phi} & 0\end{bmatrix}_{C_1}\right)\right\}\begin{bmatrix}1 & 1\\1 & -1\end{bmatrix}\right) = \frac{1}{2}\left(\begin{bmatrix}1 & 1\\1 & -1\end{bmatrix}\left\{\begin{bmatrix}0 & e^{i2\theta}\\e^{i2\Phi} & 0\end{bmatrix}_{C_0, C_1}\right\}\begin{bmatrix}1 & 1\\1 & -1\end{bmatrix}\right)$

$= \frac{1}{2}\begin{bmatrix}e^{i2\theta} + e^{i2\Phi} & -e^{i2\theta} + e^{i2\Phi}\\ e^{i2\theta} - e^{i2\Phi} & -e^{i2\theta} - e^{i2\Phi}\end{bmatrix}_{C_0, C_1}$ (10)

Where,

$\theta = \pi/4$ radians $= -\Phi$ (for RZ gates),

$\frac{1}{\sqrt{2}}\begin{bmatrix} 1 & 1 \\ 1 & -1 \end{bmatrix}$ is the unitary matrix of the H gate,

$\begin{bmatrix} 0 & 1 \\ 1 & 0 \end{bmatrix}_{C_0}$ is the unitary matrix of the CNOT gate that exists when $C_0 = |1\rangle$,

$\begin{bmatrix} 0 & 1 \\ 1 & 0 \end{bmatrix}_{C_1}$ is the unitary matrix of the CNOT gate that exists when $C_1 = |1\rangle$,

$\begin{bmatrix} 1 & 0 \\ 0 & e^{i\theta} \end{bmatrix}$ is the unitary matrix of the RZ gate, for the rotational angle $\theta$ of $(\pi/4)$ radians, and

$\begin{bmatrix} 1 & 0 \\ 0 & e^{i\Phi} \end{bmatrix}$ is the unitary matrix of the RZ gate, for the rotational angle $\Phi$ of $(-\pi/4)$ radians.

Since the target qubit (t) is initially set to the state of $|0\rangle$, then $t = \begin{bmatrix} 1 \\ 0 \end{bmatrix}$. When both controls are in the state of $|1\rangle$, then the layout-aware 3-bit Toffoli gate flips the state of target to $|1\rangle$ (by ignoring the global phase after the classical measurement), as expressed in (11) below. On the other hand, when any of the controls is in the state of $|0\rangle$, then the mirrored structure of the layout-aware 3-bit Toffoli gate collapses and the state of the target qubit remains at $|0\rangle$. Hence, the layout-aware 3-bit Toffoli gate successfully acts as a quantum counterpart of the classical Boolean 2-input AND gate.

$$\begin{aligned}
\text{Output of layout-aware 3-bit Toffoli} \atop \text{(when } C_0 = C_1 = |1\rangle \text{ and } t = |0\rangle\text{)} &= \frac{1}{2}\begin{bmatrix} e^{i2\theta} + e^{i2\Phi} & -e^{i2\theta} + e^{i2\Phi} \\ e^{i2\theta} - e^{i2\Phi} & -e^{i2\theta} - e^{i2\Phi} \end{bmatrix}_{C_0, C_1} \begin{bmatrix} 1 \\ 0 \end{bmatrix}_t \\
&= \frac{1}{2}\begin{bmatrix} e^{i2\theta} + e^{i2\Phi} \\ e^{i2\theta} - e^{i2\Phi} \end{bmatrix}_{C_0 = C_1 = |1\rangle,\ t = |0\rangle} = \begin{bmatrix} 0 \\ 1 \end{bmatrix} = |1\rangle
\end{aligned} \quad (11)$$

From (7) above, the layout-aware 4-bit Toffoli gate is the straight mirrored hierarchical structure of two cores of layout-aware 3-bit Toffoli gate and one additional control (CNOT). The layout-aware 4-bit Toffoli gate has three control qubits ($C_0$, $C_1$, and $C_2$) and one target qubit (t). Therefore, the layout-aware 4-bit Toffoli gate is mathematically modeled as expressed in (12) below.

$$\begin{aligned}
\text{Layout-aware 4-bit Toffoli gate} &= \frac{1}{\sqrt{2}}\begin{bmatrix} 1 & 1 \\ 1 & -1 \end{bmatrix} \left\{ \begin{bmatrix} 0 & e^{i2\theta} \\ e^{i2\Phi} & 0 \end{bmatrix}_{C_1, C_2} \right\} \begin{bmatrix} 0 & 1 \\ 1 & 0 \end{bmatrix}_{C_0} \left\{ \begin{bmatrix} 0 & e^{i2\theta} \\ e^{i2\Phi} & 0 \end{bmatrix}_{C_1, C_2} \right\} \frac{1}{\sqrt{2}}\begin{bmatrix} 1 & 1 \\ 1 & -1 \end{bmatrix} \\
&= \frac{1}{2}\left( \begin{bmatrix} 1 & 1 \\ 1 & -1 \end{bmatrix} \left\{ \begin{bmatrix} 0 & e^{i4\theta} \\ e^{i4\Phi} & 0 \end{bmatrix}_{C_0, C_1, C_2} \right\} \begin{bmatrix} 1 & 1 \\ 1 & -1 \end{bmatrix} \right) \\
&= \frac{1}{2}\begin{bmatrix} e^{i4\theta} + e^{i4\Phi} & -e^{i4\theta} + e^{i4\Phi} \\ e^{i4\theta} - e^{i4\Phi} & -e^{i4\theta} - e^{i4\Phi} \end{bmatrix}_{C_0, C_1, C_2}
\end{aligned} \quad (12)$$

Where,

$\theta = \pi/8$ radians $= -\Phi$ (for RZ gates),

$\begin{bmatrix} 0 & 1 \\ 1 & 0 \end{bmatrix}_{C_0}$ is the unitary matrix of the CNOT gate that exists when $C_0 = |1\rangle$, and

$\begin{bmatrix} 0 & e^{i2\theta} \\ e^{i2\Phi} & 0 \end{bmatrix}_{C_1, C_2}$ is the unitary matrix of the core of the layout-aware 3-bit Toffoli gate, which exists when both $C_1$ and $C_2$ are in the state of $|1\rangle$.

Since the target qubit is initially set to the state of $|0\rangle$, then $t = \begin{bmatrix} 1 \\ 0 \end{bmatrix}$. When all controls are in the state of $|1\rangle$, then the target qubit is flipped to the state of $|1\rangle$ (by ignoring the global phase after the classical measurement), as expressed in (13) below. On the other hand, when any of the controls is in the state of $|0\rangle$, then the mirrored structure of the layout-aware 4-bit Toffoli gate collapses and the state of the target qubit remains at $|0\rangle$. Hence, the layout-aware 4-bit Toffoli gate successfully acts as a quantum counterpart of the classical Boolean 3-input AND gate.

$$\text{Output of layout-aware 4-bit Toffoli (when } C_0 = C_1 = C_2 = |1\rangle \text{ and } t = |0\rangle\text{)} = \frac{1}{2}\begin{bmatrix} e^{i4\theta} + e^{i4\Phi} & -e^{i4\theta} + e^{i4\Phi} \\ e^{i4\theta} - e^{i4\Phi} & -e^{i4\theta} - e^{i4\Phi} \end{bmatrix}_{C_0, C_1, C_2} \begin{bmatrix} 1 \\ 0 \end{bmatrix}_t$$

$$= \frac{1}{2}\begin{bmatrix} e^{i4\theta} + e^{i4\Phi} \\ e^{i4\theta} - e^{i4\Phi} \end{bmatrix}_{C_0 = C_1 = C_2 = |1\rangle,\ t = |0\rangle} = \begin{bmatrix} 0 \\ 1 \end{bmatrix} = |1\rangle \quad (13)$$

Consequently, from (8) above, the layout-aware $n$-bit Toffoli gate is the straight mirrored hierarchical structure of two cores of layout-aware $(n-1)$-bit Toffoli gates and one additional control (CNOT), and the generic mathematical model of the layout-aware $n$-bit Toffoli gate is expressed in (14) below, where $m$ is the number of control qubits, $m = n - 1$, $n \geq 3$, and the rotational angles ($\theta$ and $\Phi$) were previously defined in (9) above.

$$\text{Layout-aware } n\text{-bit Toffoli} = \frac{1}{2}\begin{bmatrix} e^{i2^{(m-1)}\theta} + e^{i2^{(m-1)}\Phi} & -e^{i2^{(m-1)}\theta} + e^{i2^{(m-1)}\Phi} \\ e^{i2^{(m-1)}\theta} - e^{i2^{(m-1)}\Phi} & -e^{i2^{(m-1)}\theta} - e^{i2^{(m-1)}\Phi} \end{bmatrix}_{\forall C_m,\ m \geq 2} \quad (14)$$

In our research, we noticed that the layout-aware $n$-bit Toffoli gate always:

1. maintains its straight mirrored hierarchical structure through the counterclockwise and clockwise rotations of $\pm e^{i2^{(m-1)}\theta}$ and $\pm e^{i2^{(m-1)}\Phi}$, respectively, for $\theta = \pi/2^m = -\Phi$, where $m = n - 1$,
2. generates the state of $|1\rangle$ for the target qubit, when all $m$ controls are in the state of $|1\rangle$,
3. generates the state of $|0\rangle$ for the target qubit, when any of $m$ controls is in the state of $|0\rangle$, and
4. acts as the quantum counterpart of the classical Boolean $m$-input AND gate.

### 2.4 Transpilation Quantum Cost

In this paper, we introduced a new concept of calculating the quantum cost for a quantum circuit after it is transpiled into an IBM QPU, this concept is termed the "transpilation quantum cost (TQC)". The TQC is the sum of the total numbers of (i) the native single-qubit gates ($\sqrt{X}$ and RZ), (ii) the native double-qubit gates (CNOT), (iii) the number of crossing connections among the qubits, i.e., the count of SWAP gates among the qubits, and (iv) the depth (as the critical longest path through the total number of all native gates), for the final transpiled quantum circuit.

*Definition 3.* *The transpilation quantum cost (TQC) of a quantum circuit, which is transpiled into an IBM QPU, is formulated as in*

$$TQC = N_1 + N_2 + XC + D. \quad (16)$$

*Where,*
*$N_1$ is the total number of IBM native single-qubit gates,*
*$N_2$ is the total number of IBM native double-qubit gates,*
*XC is the total number of crossing connections among the physical qubits, i.e., the count of SWAP gates, and*
*D is the depth, as the critical longest path of $N_1$ and $N_2$, for the final transpiled quantum circuit □*

In the following subsection, for the layout-aware *n*-bit Toffoli gate and from (16) above, it was observed that XC = 0 for $3 \leq n \leq 4$, since XC mainly depends on the layout of an IBM QPU, i.e., how the physical qubits of an IBM QPU are arranged and connected. In our research, it was concluded that the layout-aware *n*-bit Toffoli gate always has a lower TQC than that of the conventional *n*-bit Toffoli gate, after they are transpiled using IBM QPUs.

In this paper, Maslov's quantum cost [24] is not included and compared with our proposed TQC, due to the following facts:

1. Maslov's quantum cost counts the number of conventional 3-bit Toffoli gates and CNOT gates that compose the conventional *n*-bit Toffoli gate.
2. TQC counts IBM native (single-qubit and double-qubit) gates, the crossing connections, and the depth for the layout-aware *n*-bit Toffoli gate.
3. Maslov's quantum cost is not designed for quantum circuit transpilation into an IBM QPU, since the conventional *n*-bit Toffoli gate is not constructed from IBM native gate.
4. TQC is designed for quantum circuit transpilation into an IBM QPU, since the layout-aware *n*-bit Toffoli gate is entirely constructed from IBM native gates.

### 2.5 Layouts of IBM QPUs

In general, IBM-based transpilation software (or the Transpiler [17]) includes the processing steps of compiling a quantum circuit to fit the layout of an IBM QPU. These processing steps consist of (i) decomposing the non-native gates to the native gates, (ii) mapping the utilized qubits of a quantum circuit to the physical qubits of a QPU, and (iii) re-routing the physical qubits using SWAP gates, if necessary, to match the quantum operations of a quantum circuit.

The layout defines how the physical qubits are structured, indexed, and linked inside an IBM QPU. In this paper, our design of the layout-aware *n*-bit Toffoli gate is mainly concentrated on three features: (i) the structure of layouts, (ii) the indices of physical qubits, and (iii) the direct linkages (or the direct physical channels) among physical qubits. Therefore, the layout of an IBM QPU defines how the layout-aware *n*-bit Toffoli gate is constructed as the straight mirrored hierarchical structure with a lower TQC. Note that when two adjacent physical qubits are connected together through a direct linkage (or a direct channel), then the SWAP gate is not required.

Conversely, other parameters and specifications of IBM QPUs, including quantum volume (QV), circuit layer operations per second (CLOPS), physical qubit's details (T1, T2, frequency, and anharmonicity), and median native gate's errors ($\sqrt{X}$, CNOT, and readout) [25-27], are not considered for the purpose of our research in designing the layout-aware *n*-bit Toffoli gate.

In our research, we investigated the following IBM QPUs from the Open Plan of the IBM quantum system [28]. These IBM QPUs are fabricated in three different layouts for a defined number of physical qubits:

1. The linear layout of five physical qubits for the QPU (`ibmq_manila`), and Figure 8 (a) depicts the schematic of this layout with its qubits' indices and linkages.
2. The T-like layout of five physical qubits for the QPUs (`ibmq_lima`, `ibmq_belem`, and `ibmq_quito`), and Figure 8 (b) illustrates the schematic of this layout with its qubits' indices and linkages.
3. The I-like layout of seven physical qubits for the QPUs (`ibmq_jakarta`, `ibmq_nairobi`, `ibmq_lagos`, and `ibmq_perth`), and Figure 8 (c) demonstrates the schematic of this layout with its qubits' indices and linkages.

In our experiments, we only utilized three IBM QPUs, which are `ibmq_manila`, `ibmq_quito`, and `ibmq_perth`, for different layouts and different numbers of physical qubits.

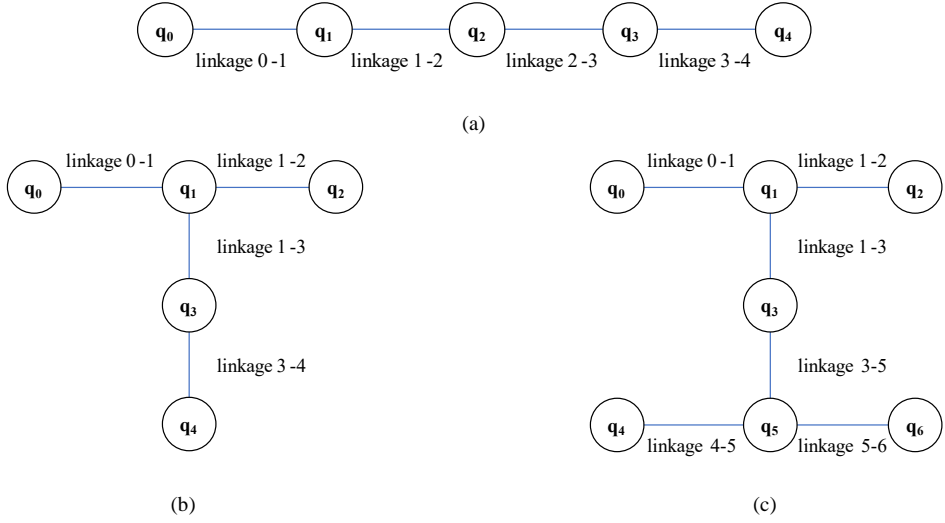

Figure 8: Schematics of different layouts of IBM QPUs with indexed physical qubits (q) and linkages (in blue): (a) linear layout of five physical qubits and four linkages for the QPU (`ibmq_manila`), (b) T-like layout of five physical qubits and four linkages for the QPUs (`ibmq_lima`, `ibmq_belem`, and `ibmq_quito`), and (c) I-like layout of seven physical qubits with six linkages for the QPUs (`ibmq_jakarta`, `ibmq_nairobi`, `ibmq_lagos`, and `ibmq_perth`).

In Figure 8 (a), (b), and (c), the (linkage 0-1) is the direct communication channel between the physical qubits ($q_0$ and $q_1$), and the SWAP gate is not required. However, when the ($q_0$) needs to communicate with the ($q_2$), then the re-routing process is required using two SWAP gates: (i) the first SWAP gate switches ($q_1$) and ($q_2$), so that ($q_0$) can now communicate with ($q_2$), and (ii) the second SWAP gate re-switches ($q_2$) and ($q_1$) for indices resetting. A single non-native SWAP gate can be decomposed into three native CNOT gates, as shown in Figure 9.

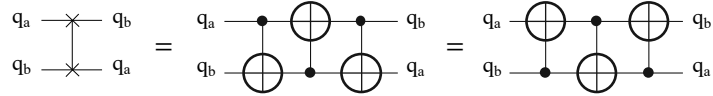

Figure 9: Schematics of the non-native SWAP gate (left-side) and its equivalent decomposition of three native CNOT (Feynman) gates (right-side), for swapping arbitrary indexed physical qubits ($q_a$ and $q_b$) of an IBM QPU.

For the layout-aware $n$-bit Toffoli gate, to maintain its straight mirrored hierarchical structure, the target qubit is mapped to a physical qubit that has direct linkages with other physical qubits, and these other physical qubits are the ($n$–1) controls. Such a target mapping ensures that the re-routing process is not required during the transpilation, i.e., no SWAP gates are added among the controls. Thus, the TQC will be minimized for the final transpiled quantum circuit, since XC = 0.

For an IBM QPU, the layout and the number of $n$ physical qubits play two crucial roles in: (i) minimizing the TQC, and (ii) maintaining the straight mirrored hierarchical structure for the layout-aware $n$-bit Toffoli gate. For that, in this paper, we proposed the following two parameters when constructing the layout-aware $n$-bit Toffoli gate:

1. The "optimal-$n$" parameter defines the neighboring number of $n$ physical qubits, where $3 \leq n \leq 4$, for any three (or four) adjacent physical qubits, which are directly connected within two (or three) linkages.
2. The "critical-$n$" parameter defines the arbitrary number of $n$ physical qubits, where $3 \leq n \leq$ total number of physical qubits, for arbitrary physical qubits that are not (or mostly not) directly connected within linkages.

The optimal-*n* parameter ensures to obtain the aforementioned two crucial roles (minimizing the TQC and maintaining the structure). The critical-*n* parameter causes the layout-aware *n*-bit Toffoli gate to lose its straight mirrored hierarchical structure, due to the added SWAP gates among the physical qubits; thereafter, the TQC increases. Table 2 states these two crucial roles with their two associated parameters.

Table 2: Layouts and required *n* physical qubits (as the crucial roles) for IBM QPUs, and their associated optimal-*n* and critical-*n* (as the parameters) for the layout-aware *n*-bit Toffoli gate.

| IBM QPUs | | | | Required *n* physical qubits for the layout-aware *n*-bit Toffoli gate | |
|---|---|---|---|---|---|
| Codename | Layout | Total number of physical qubits | Total number of linkages | Optimal-*n* | Critical-*n* |
| ibmq_manila | Linear | 5 | 4 | 3 | 4 or 5 |
| ibmq_quito | T-like | 5 | 4 | 3 or 4 | 4 or 5 |
| ibmq_perth | I-like | 7 | 6 | 3 or 4 | 4, 5, 6, or 7 |

***Definition 4.*** *The optimal-n is the parameter that defines the neighboring number of n physical qubits, to construct the layout-aware n-bit Toffoli gate for an IBM QPU of the layouts (linear, T-like, and I-like), where $3 \leq n \leq 4$. The optimal-n parameter ensures to (i) have any three (or four) adjacent physical qubits being directly connected through two (or three) linkages, (ii) minimize the transpilation quantum cost (TQC), and (iii) maintain the straight mirrored hierarchical structure for the final transpiled quantum circuit* □

***Definition 5.*** *The critical-n is the parameter that defines the arbitrary number of n physical qubits, to construct the layout-aware n-bit Toffoli gate for an IBM QPU of the layouts (linear, T-like, and I-like), where $3 \leq n \leq$ total number of physical qubits. From the critical-n parameter, (i) arbitrary physical qubits are not (or mostly not) directly connected within linkages, (ii) the layout-aware n-bit Toffoli gate loses its straight mirrored hierarchical structure, (iii) SWAP gates are added among the physical qubits, and (iv) the transpilation quantum cost (TQC) is increased as n increases* □

For all layouts (linear, T-like, and I-like) of an IBM QPU, our proposed parameters (optimal-*n* and critical-*n*) are proportionally related to the minimum and maximum of TQC, as in

$$\text{Parameters (optimal-}n\text{ and critical-}n\text{)} \propto \text{TQC}, \tag{17}$$

$$\min_{\text{optimal-}n} \text{TQC}, \tag{18}$$

$$\max_{\text{critical-}n} \text{TQC}. \tag{19}$$

From Table 2 and (17-19) above, the optimal-*n* acts as the thresholding parameter (in selecting *n* physical qubits) between the minimum TQC (when $3 \leq n \leq 4$) and the maximum TQC (when $n > 4$). For the critical-*n* parameter, when *n* increases for a specific layout, more re-routing processes are performed, i.e., more SWAP gates are added to the final transpiled quantum circuit. Thus, the TQC increases as well, due to the enlargement of $N_2$, XC, and D, as previously stated in (16) above.

For the optimal-*n* parameter, different *n* physical qubits can be configured (or chosen) based on the different layouts of IBM QPUs, and all these configurations are selected to obtain the aforementioned two crucial roles. For *n* = 3, different configurations for the optimal-*n* parameter can be utilized. Figure 10 depicts three configurations for the linear layout, while Figure 11 illustrates four configurations for the T-like layout, and Figure 12 demonstrates seven configurations for the I-like layout; see Table 2.

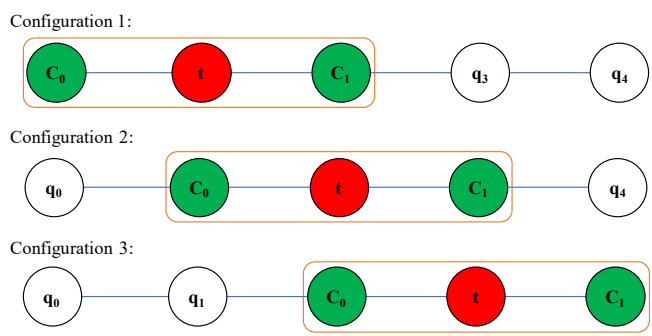

Figure 10: Schematics of three different configurations for the optimal-$n$ parameter, where $n = 3$, for the linear layout of `ibmq_manila` QPU. Note that the orange area represents a single configuration, the linkages are in blue lines, the green circles denote the two controls ($C_0$ and $C_1$), and the red circle identifies the target (t) for the layout-aware 3-bit Toffoli gate.

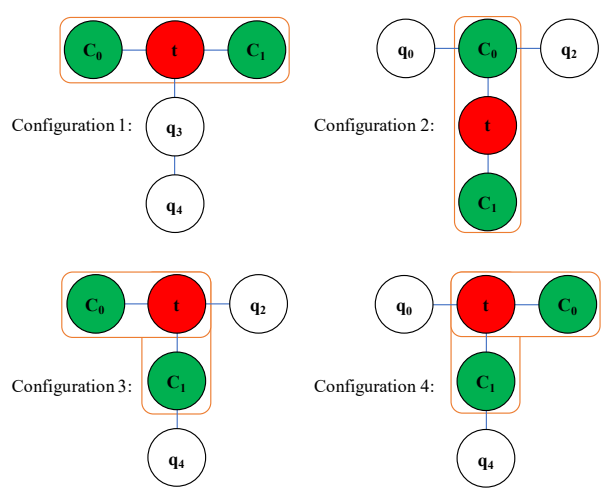

Figure 11: Schematics of four different configurations for the optimal-$n$ parameter, where $n = 3$, for the T-like layout of `ibmq_quito` QPU. Note that the orange area represents a single configuration, the linkages are in blue lines, the green circles denote the two controls ($C_0$ and $C_1$), and the red circle identifies the target (t) for the layout-aware 3-bit Toffoli gate.

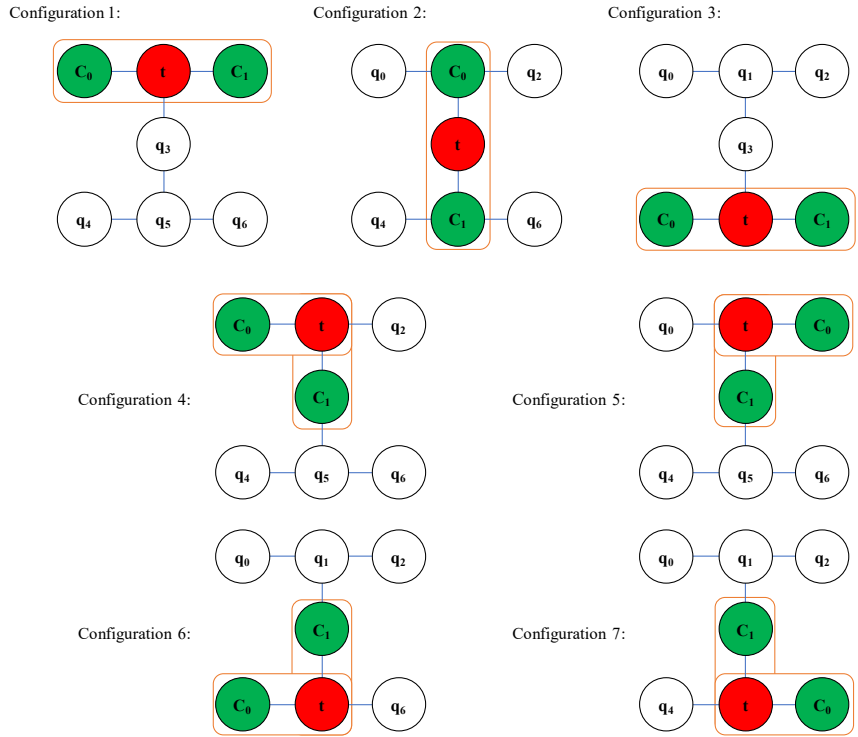

Figure 12: Schematics of seven different configurations for the optimal-$n$ parameter, where $n = 3$, for the I-like layout of ibmq_perth QPU. Note that the orange area represents a single configuration, the linkages are in blue lines, the green circles denote the two controls ($C_0$ and $C_1$), and the red circle identifies the target (t) for the layout-aware 3-bit Toffoli gate.

For $n = 4$, different configurations for the optimal-$n$ parameter can be utilized. Figure 13 illustrates the single configuration for the T-like layout, and Figure 14 demonstrates two different configurations for the I-like layout. The optimal-$n$ parameter (where $n = 4$) cannot be utilized for the linear layout of ibmq_manila QPU, due to the added SWAP gates among the three control qubits; thus, the final TQC increases. Therefore, the critical-$n$ will be the parameter for the linear layout, when $n = 4$; see Table 2.

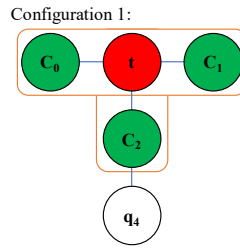

Figure 13: Schematic of single configuration for the optimal-$n$ parameter, where $n = 4$, for the T-like layout of ibmq_quito QPU. Note that the orange area represents this single configuration, the linkages are in blue lines, the green circles denote the three controls ($C_0$, $C_1$, and $C_2$), and the red circle identifies the target (t) for the layout-aware 4-bit Toffoli gate.

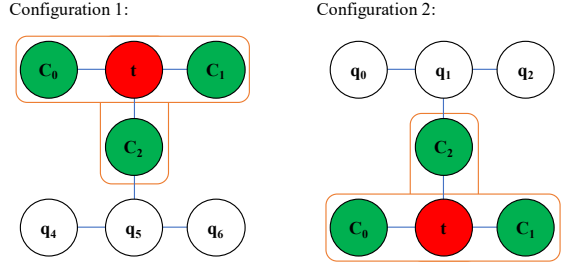

Figure 14: Schematics of two configurations for the optimal-$n$ parameter, where $n = 4$, for the I-like layout of `ibmq_perth` QPU. Note that the orange area represents a single configuration, the linkages are in blue lines, the green circles denote the three controls ($C_0$, $C_1$, and $C_2$), and the red circle identifies the target (t) for the layout-aware 4-bit Toffoli gate.

For $n > 4$, all configurations of the optimal-$n$ parameters cannot be utilized for all layouts (linear, T-like, and I-like) of an IBM QPU, due to the added SWAP gates among the four (or more) control qubits; thus, the final TQC will dramatically increase. Therefore, arbitrary configurations can be implemented when $n > 4$, i.e., arbitrary configurations for the critical-$n$ parameter; see Table 2.

When $3 \leq n \leq 4$, the control qubits ($C_0$, $C_1$, and/or $C_2$) can replace each other, based on the design and function of a quantum circuit. For instance, from "Configuration 1" in Figure 10, the physical qubit ($q_0$) can be utilized for the control ($C_1$) instead of the control ($C_0$), and the physical qubit ($q_2$) can be used for the control ($C_0$) instead of the control ($C_1$); in such a case, another (fourth) configuration can be implemented. To construct a single layout-aware $n$-bit Toffoli gate (of $n-1$ controls, one target, and $3 \leq n \leq 4$), Table 3 states the maximum number of configurations for the optimal-$n$ parameter, for all layouts (linear, T-like, and I-like) of IBM QPUs. Note that N.A. means "not applicable".

Table 3: Maximum number of configurations for all layouts (linear, T-like, and I-like) of IBM QPUs, for constructing a single layout-aware $n$-bit Toffoli gate, where $3 \leq n \leq 4$.

| IBM QPUs | | | | Single layout-aware $n$-bit Toffoli gate, |
|---|---|---|---|---|
| **Codename** | **Layout** | **$n$ utilized physical qubits** | **Number of direct linkages** | **Maximum number of configurations (the optimal-$n$ parameter)** |
| ibmq_manila | Linear | 3 | 2 | 6 |
|  |  | 4 | N.A. | N.A. |
| ibmq_quito | T-like | 3 | 2 | 8 |
|  |  | 4 | 3 | 6 |
| ibmq_perth | I-like | 3 | 2 | 14 |
|  |  | 4 | 3 | 12 |

From Table 3, the maximum number of configurations gives the freedom of designing the layout-aware $n$-bit Toffoli gates, where $3 \leq n \leq 4$, for an IBM QPU. This freedom of design utilizes the most adjacent physical qubits through their direct linkages, and it leaves the other non-directly connected physical qubits for other quantum operations or gates. All configurations for the optimal-$n$ parameter ensure obtaining the two crucial roles (minimizing the TQC and maintaining the structure) for the layout-aware $n$-bit Toffoli gate, where $3 \leq n \leq 4$.

Finally, our experiments were performed on all layouts (linear, T-like, and I-like) for IBM QPUs (`ibmq_manila`, `ibmq_quito`, and `ibmq_perth`). However, when IBM quantum system (as the quantum company provider) proposes a new layout, e.g., hexagonal or mesh-like, in the future, then:

- If the neighboring physical qubits are in linear, T-like, or I-like connection, then the optimal-$n$ parameter is preserved for $3 \leq n \leq 4$.
- If the neighboring physical qubits are in a Plus-like (+) or Multiply-like (×) connection, then the optimal-$n$ parameter will be in the range of $3 \leq n \leq 5$.
- If the neighboring physical qubits are in an Asterisk-like (✱) connection, then the optimal-$n$ parameter will be in the range of $3 \leq n \leq 7$, and so on for different geometrical connections.

### 2.6 Computer Code

In our experiments, Python language and Qiskit (as the quantum computing toolkit from IBM Quantum Lab [29, 30]) are employed to design, examine, and evaluate the layout-aware $n$-bit Toffoli gate for different layouts of IBM QPUs. In this paper, the Python-Qiskit function, called `laToffoli()`, constructs the layout-aware $n$-bit Toffoli gate for a specific layout of an IBM QPU. The parameters (`controls` and `target`) of the `laToffoli()` function define such a specific layout, by using the indices of physical qubits of an IBM QPU.

For transpiling the layout-aware $n$-bit Toffoli gate into an IBM QPU:

1. The `laToffoli()` function is our proposed transpilation software package, and the IBM-based transpilation software (Transpiler) is not fully utilized in the decomposition process, since all quantum gates are native gates.
2. The Transpiler is utilized for mapping the qubits of a quantum circuit to the physical qubits.
3. The Transpiler is utilized, if required, for re-routing the physical qubits, by adding SWAP gates.

In general, the Qiskit-based quantum instructions of the laToffoli() function are based on:

1. The workflow of ALGORITHM 1.
2. The XY-plane of the Bloch sphere visualization for the core of layout-aware 3-bit Toffoli gate, as demonstrated in Figure 4.
3. The cores of layout-aware 3-bit Toffoli gate and layout-aware $n$-bit Toffoli gate, as previously stated in (5) and (8) above, respectively.
4. The quantum circuits for the layout-aware 3-bit Toffoli gate and the layout-aware $n$-bit Toffoli gate, as illustrated in Figure 3 and Figure 7, respectively.

```
def laToffoli(controls, target):
    """
    Where,
    laToffoli means layout-aware Toffoli.
    controls: list of indices for m controls (physical input qubits).
    target:   index of the target (physical output qubit).
    Returns:  the quantum circuit of layout-aware n-bit Toffoli gate.
    """
```

```python
m = len(controls)    # Total number of m controls
n = m + 1            # Total number of all qubits (m controls + 1 target)

# Rotational angles in the XY-plane of the Bloch sphere:
π = np.pi            # Do not forget to "import numpy as np"
θ = π/(2**m)         # Counterclockwise rotational angle for RZ
Φ = -θ               # Clockwise rotational angle for RZ

toffoli_gate = QuantumCircuit(n)   # Constructing the core of layout-aware 3-bit Toffoli
toffoli_gate.rz(θ, target)
toffoli_gate.cx(controls[-1], target)
toffoli_gate.rz(Φ, target)
toffoli_gate.cx(controls[-2], target)
toffoli_gate.rz(θ, target)
toffoli_gate.cx(controls[-1], target)
toffoli_gate.rz(Φ, target)

r = m – 2            # The remaining m controls:

# Constructing the cores of layout-aware r-bit Toffoli:
while( r > 0 ):
    r = r - 1
    backup = toffoli_gate.copy()
    toffoli_gate.cx(controls[r], target)
    toffoli_gate.compose(backup, inplace=True, wrap=False)

qc = QuantumCircuit(n)      # Constructing the completed layout-aware n-bit Toffoli
qc.h(target)                # Adding the first H gate
qc.compose(toffoli_gate, inplace=True, wrap=False)    # Adding the core
qc.h(target)                # Adding the second H gate

return qc                   # Returning the layout-aware n-bit Toffoli
```

## 3 CASE STUDIES AND RESULTS

Different experiments of the conventional $n$-bit Toffoli gates and the layout-aware $n$-bit Toffoli gates are transpiled, from using IBM QPUs of different layouts and the number of physical qubits. In this paper, we utilized the following IBM QPUs: (i) the `ibmq_manila` of linear layout for five physical qubits, (ii) the `ibmq_quito` of T-like layout for five physical qubits, and (iii) the `ibmq_perth` of I-like layout for seven physical qubits.

These experiments are transpiled and compared in the manner of the transpilation quantum cost (TQC) and the parameters (optimal-$n$ and critical-$n$), for $3 \leq n \leq 7$ physical qubits. After transpilation, we observed that the transpiled layout-aware $n$-bit Toffoli gate (by using our `laToffoli()` transpilation package) always has a lower TQC than that of the transpiled conventional $n$-bit Toffoli gate (by using the IBM-based transpilation software, Transpiler).

The IBM Quantum Lab was utilized to design and examine these experiments, for 1024 shots [31, 32]. However, the specifications of QV, CLOPS, T1, T2, frequency, anharmonicity, and median errors of the native gate, speed of transpilation, speed of Internet connection, usage load of an IBM QPU are not considered for the purpose of our research and experiments.

For ease of illustration and experimentation, the constructions of layout-aware $n$-bit Toffoli gate are based on "Configuration 1", as previously demonstrated in Figure 10, Figure 11, Figure 12, Figure 13, and Figure 14. In contrast, the conventional $n$-bit Toffoli gate is constructed using the "standard approach". The standard approach means that the $n-1$ controls occupy the first in-sequence physical qubits, and the target occupies the following physical qubit after the last control. Note that, for the standard approach, SWAP gates, i.e., the crossing connections among the physical qubits, always exist.

### 3.1 Experiments for $n$ = 3

The conventional 3-bit Toffoli gate is constructed using the standard approach, and the layout-aware 3-bit Toffoli gate is constructed using Configuration 1, as shown in Figure 15 (a) and Figure 15 (b), respectively. Both 3-bit Toffoli gates are then transpiled into three IBM QPUs (`ibmq_manila`, `ibmq_quito`, and `ibmq_perth`). After transpilation: (i) the controls and target of conventional 3-bit Toffoli gate are randomly mapped to any physical qubits for all QPUs, as illustrated in Figure 16, and (ii) the controls and target of layout-aware 3-bit Toffoli gate are exactly mapped to the same indexed physical qubits for all QPUs, as demonstrated in Figure 17, i.e., optimal-$n$ = 3. The status of physical qubits mapping for all IBM QPUs is expressed in Table 5.

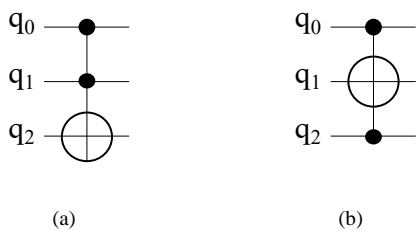

(a)          (b)

Figure 15: Schematics for the construction of: (a) the conventional 3-bit Toffoli gate for the target qubit ($q_2$), and (b) the layout-aware 3-bit Toffoli gate for the target qubit ($q_1$). Note the indices of qubits for controls and targets of both 3-bit Toffoli gates.

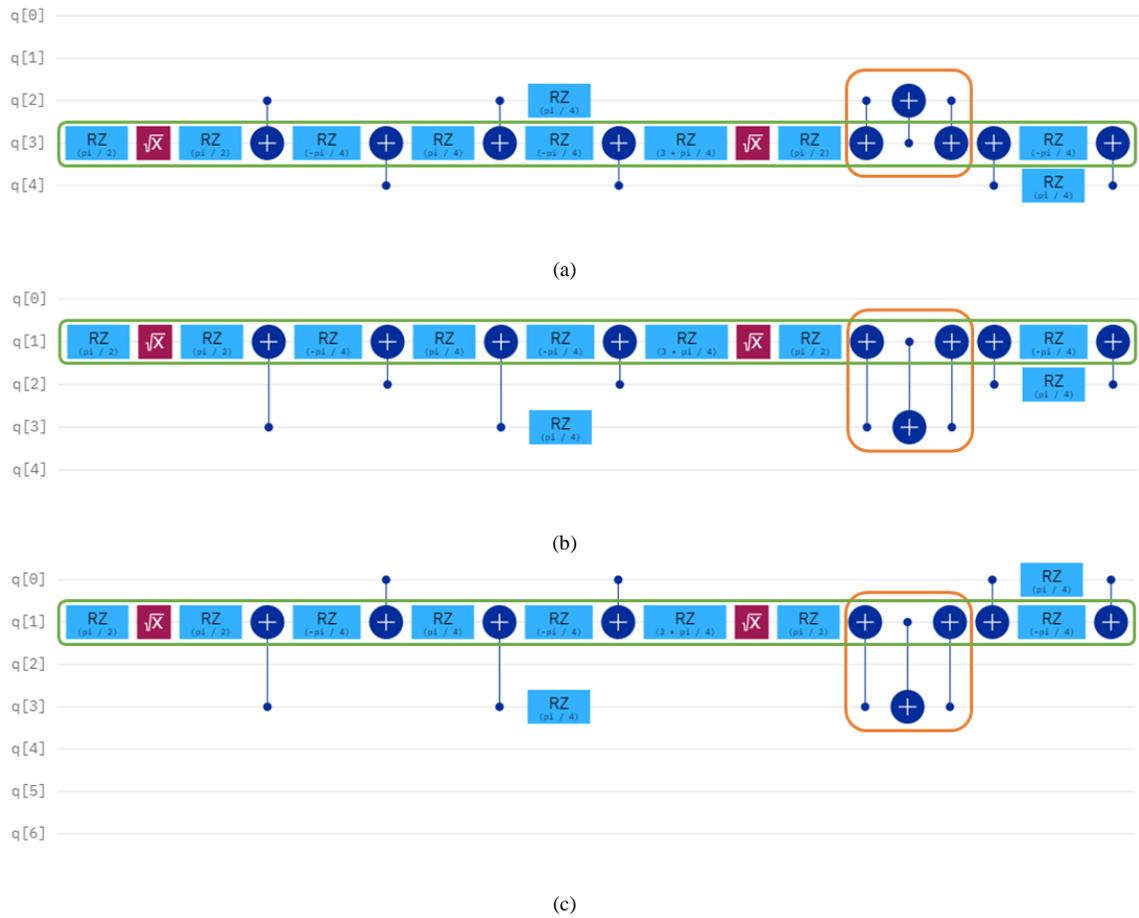

Figure 16: The transpiled conventional 3-bit Toffoli gates for: (a) `ibmq_manila` QPU, (b) `ibmq_quito` QPU, and (c) `ibmq_perth` QPU. Note that the orange area represents a single SWAP gate (XC), the green area denotes the total depth (D), and the controls and target of the conventional 3-bit Toffoli gate are randomly mapped to any physical qubits for all QPUs. For all transpiled quantum circuits, $N_1 = 12$, $N_2 = 9$, XC = 1, and D = 29, so TQC = 41.

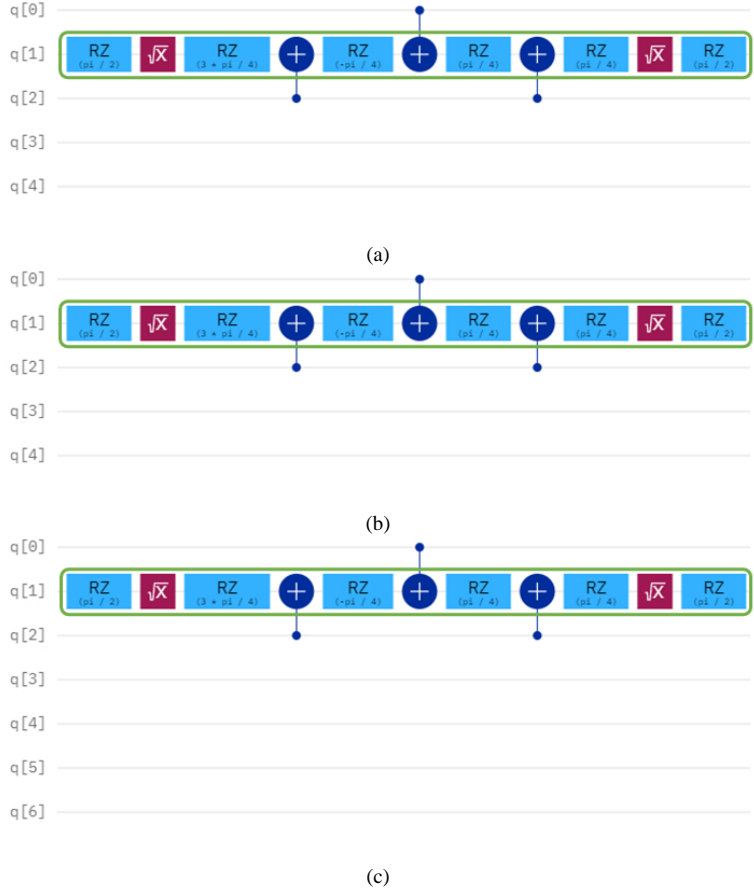

(a)

(b)

(c)

Figure 17: The transpiled layout-aware 3-bit Toffoli gates for: (a) ibmq_manila QPU, (b) ibmq_quito QPU, and (c) ibmq_perth QPU. Note that there are no SWAP gates, the green area denotes the total depth (D), and the controls and target of the layout-aware 3-bit Toffoli gate are exactly mapped to the same indexed physical qubits for all QPUs. For all transpiled quantum circuits, $N_1 = 8$, $N_2 = 3$, XC = 0, and D = 11, so TQC = 22.

## 3.2 Experiments for *n* = 4

The conventional 4-bit Toffoli gate is constructed using the standard approach, and the layout-aware 4-bit Toffoli gate is constructed using Configuration 1, as shown in Figure 18 (a) and Figure 18 (b), respectively. Both 4-bit Toffoli gates are then transpiled into three IBM QPUs (ibmq_manila, ibmq_quito, and ibmq_perth). After transpilation: (i) the controls and target of conventional 4-bit Toffoli gate are randomly mapped to any physical qubits for all QPUs, (ii) the controls and target of layout-aware 4-bit Toffoli gate are randomly mapped to any physical qubits for ibmq_manila QPU, i.e., critical-*n* = 4, and (iii) the controls and target of layout-aware 4-bit Toffoli gate are exactly mapped to the same indexed physical qubits for ibmq_quito and ibmq_perth QPUs, i.e., optimal-*n* = 4. The transpiled quantum circuits are not illustrated in this subsection, to save space. The final TQCs for both 4-bit Toffoli gates are stated in Table 4, and the status of physical qubits mapping for all IBM QPUs is expressed in Table 5.

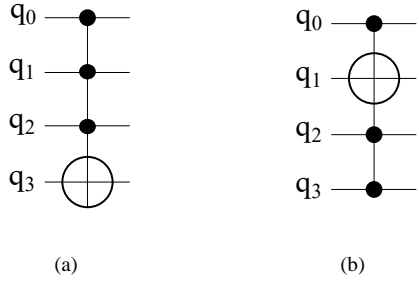

(a)            (b)

Figure 18: Schematics for the construction of: (a) the conventional 4-bit Toffoli gate for the target qubit ($q_3$), and (b) the layout-aware 4-bit Toffoli gate for the target qubit ($q_1$). Note the indices of qubits for controls and targets of both 4-bit Toffoli gates.

### 3.3 Experiments for *n* = 5

The conventional 5-bit Toffoli gate is constructed using the standard approach, and the layout-aware 5-bit Toffoli gate is constructed using Configuration 1, as shown in Figure 19 (a) and Figure 19 (b), respectively. Both 5-bit Toffoli gates are then transpiled into three IBM QPUs (`ibmq_manila`, `ibmq_quito`, and `ibmq_perth`). After transpilation: (i) the controls and target of conventional 5-bit Toffoli gate are randomly mapped to any physical qubits for all QPUs, and (ii) the controls and target of layout-aware 5-bit Toffoli gate are randomly mapped to any physical qubits for all QPUs, i.e., critical-*n* = 5. The transpiled quantum circuits are not illustrated in this subsection, to save space. The final TQCs for both 5-bit Toffoli gates are stated in Table 4, and the status of physical qubits mapping for all IBM QPUs is expressed in Table 5.

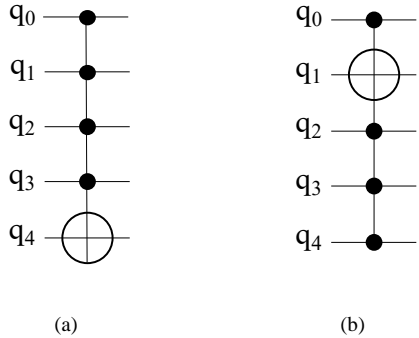

(a)            (b)

Figure 19: Schematics for the construction of: (a) the conventional 5-bit Toffoli gate for the target qubit ($q_4$), and (b) the layout-aware 5-bit Toffoli gate for the target qubit ($q_1$). Note the indices of qubits for controls and targets of both 5-bit Toffoli gates.

### 3.4 Experiments for *n* = 6

The conventional 6-bit Toffoli gate is constructed using the standard approach, and the layout-aware 6-bit Toffoli gate is constructed using Configuration 1, as shown in Figure 20 (a) and Figure 20 (b), respectively. Both 6-bit Toffoli gates are then transpiled into one IBM QPU (`ibmq_perth`). After transpilation: (i) the controls and target of conventional 6-bit Toffoli gate are randomly mapped to any physical qubits for this QPU, and (ii) the controls and target of layout-aware 6-bit Toffoli gate are randomly mapped to any physical qubits for this QPU, i.e., critical-*n* = 6. The transpiled quantum circuits are not illustrated in this subsection, to save space. The final TQCs for both 6-bit Toffoli gates are stated in Table 4, and the status of physical qubits mapping for this IBM QPU is expressed in Table 5.

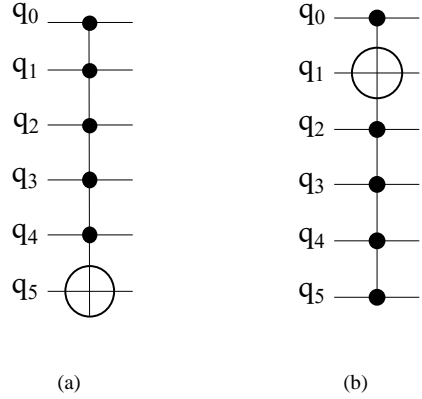

(a)　　　　　　　　(b)

Figure 20: Schematics for the construction of: (a) the conventional 6-bit Toffoli gate for the target qubit ($q_5$), and (b) the layout-aware 6-bit Toffoli gate for the target qubit ($q_1$). Note the indices of qubits for controls and targets of both 6-bit Toffoli gates.

### 3.5 Experiments for $n = 7$

The conventional 7-bit Toffoli gate and the layout-aware 7-bit Toffoli gate are constructed using the standard approach, as shown in Figure 21 (a) and Figure 21 (b), respectively. Both 7-bit Toffoli gates are then transpiled into one IBM QPU (`ibmq_perth`). After transpilation: (i) the controls and target of conventional 7-bit Toffoli gate are randomly mapped to any physical qubits for this QPU, and (ii) the controls and target of layout-aware 7-bit Toffoli gate are randomly mapped to any physical qubits for this QPU, i.e., critical-$n = 7$. The transpiled quantum circuits are not illustrated in this subsection, to save space. The final TQCs for both 7-bit Toffoli gates are stated in Table 4, and the status of physical qubits mapping for this IBM QPU is expressed in Table 5.

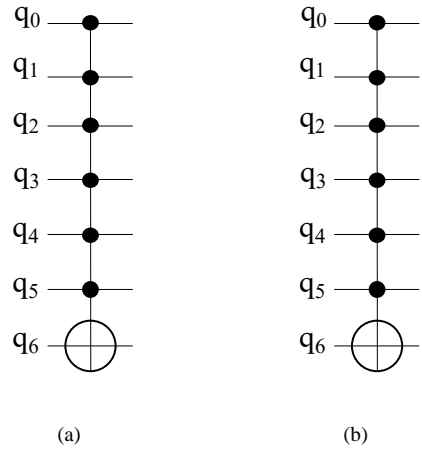

(a)　　　　　　　　(b)

Figure 21: Schematics for the construction of: (a) the conventional 7-bit Toffoli gate for the target qubit ($q_6$), and (b) the layout-aware 7-bit Toffoli gate for the target qubit ($q_6$). Note the indices of qubits for controls and targets of both 7-bit Toffoli gates.

Table 4 summarized our proposed transpilation quantum cost (TQC), as earlier discussed in (16) above, for the final transpiled quantum circuits of the aforementioned experiments (for the conventional $n$-bit Toffoli gate as compared to the layout-aware $n$-bit Toffoli gate), using different layouts and $n$ physical qubits of IBM QPUs, where $3 \leq n \leq 7$.

Table 4: Summary of TQC for the transpiled $n$-bit Toffoli gates (conventional $n$-bit Toffoli gate and layout-aware $n$-bit Toffoli gate) using IBM QPUs, where $3 \leq n \leq 7$.

| IBM QPUs of different layouts and $n$ utilized physical qubits | | | Conventional $n$-bit Toffoli gate (IBM-based transpilation) | | | | | Layout-aware $n$-bit Toffoli gate (Our transpilation software) | | | | |
|---|---|---|---|---|---|---|---|---|---|---|---|---|
| Codename | Layout | $n$ | $N_1$ | $N_2$ | XC | D | TQC | $N_1$ | $N_2$ | XC | D | TQC |
| ibmq_manila | Linear | | 12 | 9 | 1 | 19 | **41** | 8 | 3 | 0 | 11 | **22** |
| ibmq_quito | T-like | 3 | 12 | 9 | 1 | 19 | **41** | 8 | 3 | 0 | 11 | **22** |
| ibmq_perth | I-like | | 12 | 9 | 1 | 19 | **41** | 8 | 3 | 0 | 11 | **22** |
| ibmq_manila | Linear | | 20 | 29 | 5 | 30 | **84** | 12 | 13 | 2 | 19 | **46** |
| ibmq_quito | T-like | 4 | 20 | 20 | 2 | 33 | **75** | 12 | 7 | 0 | 19 | **38** |
| ibmq_perth | I-like | | 20 | 29 | 5 | 30 | **84** | 12 | 7 | 0 | 19 | **38** |
| ibmq_manila | Linear | | 61 | 87 | 17 | 77 | **242** | 20 | 33 | 6 | 41 | **100** |
| ibmq_quito | T-like | 5 | 61 | 69 | 11 | 51 | **192** | 20 | 21 | 2 | 35 | **78** |
| ibmq_perth | I-like | | 61 | 69 | 11 | 51 | **192** | 20 | 21 | 2 | 35 | **78** |
| ibmq_perth | I-like | 6 | 98 | 173 | 26 | 159 | **456** | 36 | 55 | 8 | 76 | **175** |
| ibmq_perth | I-like | 7 | 194 | 335 | 49 | 284 | **862** | 68 | 123 | 20 | 152 | **363** |

From Table 4, it is concluded that the layout-aware $n$-bit Toffoli gate can be employed as the transpilation library for the IBM quantum system, since its transpiled quantum circuit always has a much lower TQC than that of the transpiled conventional $n$-bit Toffoli gate, where $3 \leq n \leq 7$. It is also observed that the TQC for the conventional $n$-bit Toffoli gate is much higher than that of the layout-aware $n$-bit Toffoli gate, as approximately twice higher as: (i) 180% to 220% for $3 \leq n \leq 4$, and (ii) 240% to 260% for $5 \leq n \leq 7$.

In our research, we observed that the indices of $n$ qubits of Toffoli gates (the conventional $n$-bit Toffoli gate and the layout-aware $n$-bit Toffoli gate) are mapped either "exactly" or "randomly" to the indices of $n$ physical qubits of IBM QPUs. Table 5 summarizes the mapping status from the $n$ qubits of the aforementioned experiments to the $n$ physical qubits of the aforementioned IBM QPUs, where $3 \leq n \leq 7$. Note that E means "exactly mapped", R means "randomly mapped", and N.A. means "not applicable".

Table 5: Status of mapping from the $n$ qubits of transpiled Toffoli gates (conventional $n$-bit Toffoli gate and layout-aware $n$-bit Toffoli gate) to the $n$ physical qubits of IBM QPUs, where $3 \leq n \leq 7$.

| IBM QPUs | | Conventional $n$-bit Toffoli gates | | | | | Layout-aware $n$-bit Toffoli gates | | | | |
|---|---|---|---|---|---|---|---|---|---|---|---|
| Codename | Layout | $n$=3 | $n$=4 | $n$=5 | $n$=6 | $n$=7 | $n$=3 | $n$=4 | $n$=5 | $n$=6 | $n$=7 |
| ibmq_manila | Linear | R | R | R | N.A. | | E | R | R | N.A. | |
| ibmq_quito | T-like | R | R | R | | | E | E | R | | |
| ibmq_perth | I-like | R | R | R | R | R | E | E | R | R | R |

From Table 2, Table 4, and Table 5:

- For the conventional *n*-bit Toffoli gate (3 ≤ *n* ≤ 7), the utilized physical qubits of all QPUs are randomly mapped, and more re-routing processes are performed, i.e., more SWAP gates are added. This is why $N_2$, XC, and D are increased as *n* increases, which leads to TQC dramatically increases.
- For the layout-aware *n*-bit Toffoli gate (*n* = 3), the utilized physical qubits of all QPUs are exactly mapped, i.e., optimal-*n* = 3; therefore, no re-routing processes are performed, i.e., XC = 0.
- For the layout-aware *n*-bit Toffoli gate (*n* = 4), the utilized physical qubits of ibmq_quito and ibmq_perth QPUs are exactly mapped, i.e., optimal-*n* = 4; therefore, no re-routing processes are performed, i.e., XC = 0. However, the utilized physical qubits of ibmq_manila QPU are randomly mapped, and this is why critical-*n* = 4 and XC = 2.
- For the layout-aware *n*-bit Toffoli gate (5 ≤ *n* ≤ 7), the utilized physical qubits of all QPUs are randomly mapped, and this is why 5 ≤ critical-*n* ≤ 7 and XC is increased as *n* increases; therefore, TQC increases as well.

For all transpiled layout-aware *n*-bit Toffoli gates, Figure 22 demonstrates the proportional relationships between the TQCs and the two parameters (optimal-*n* and critical-*n*), using IBM QPUs (ibmq_manila, ibmq_quito, and ibmq_perth), where 3 ≤ *n* ≤ 7. Such that, the TQCs are increased as these two parameters increase. In a general aspect, *n* of 4 is considered as the separation barrier for lower and higher values of optimal-*n* and critical-*n* (see Table 2), and for TQCs (see Table 4).

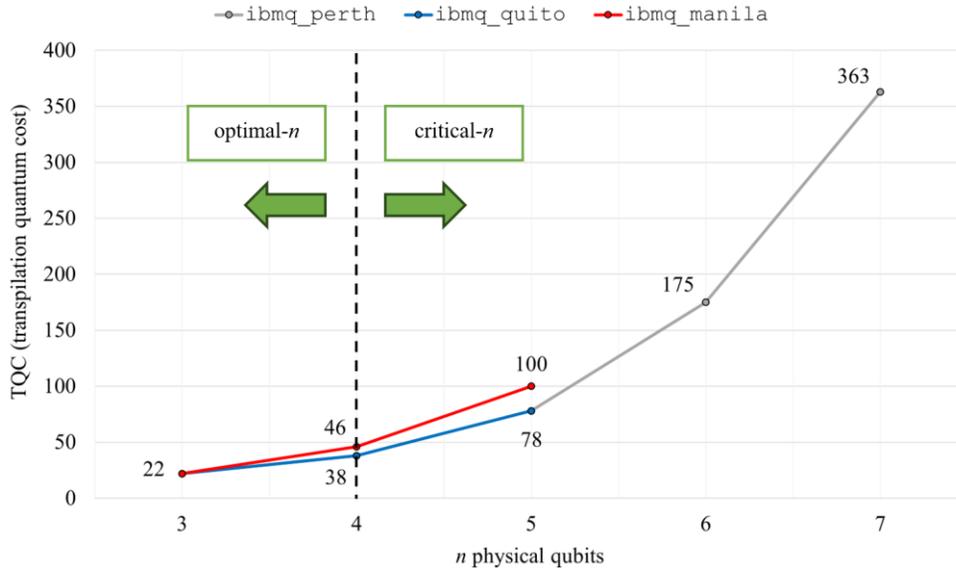

Figure 22: The proportional relationships between the TQCs and the parameters (optimal-*n* and critical-*n*), after transpiling the layout-aware *n*-bit Toffoli gates using IBM QPUs (ibmq_manila in red, ibmq_quito in blue, and ibmq_perth in gray), where 3 ≤ *n* ≤ 7. The TQCs are decreased/increased as the parameters (optimal-*n* and the critical-*n* in the green arrows) decrease/increase, respectively. For *n* = 4, these proportional relationships are separated, in a general aspect, into two regions (lower and higher values).

## 4 CONCLUSION

The conventional *n*-bit Toffoli gate is a non-native quantum gate for IBM quantum processing units (QPUs). For that, it is decomposed into a set of native quantum gates to fit the layout (or topology) of an IBM QPU. In IBM terminologies, the decomposition and fitting mechanisms belong to the transpilation process. In our research of transpiling the conventional *n*-bit Toffoli gate, the following IBM QPUs were utilized: (i) `ibmq_manila` of linear layout for five physical qubits, (ii) `ibmq_quito` of T-like layout for five physical qubits, and (iii) `ibmq_perth` of I-like layout for seven physical qubits. After transpiling the conventional *n*-bit Toffoli gate, for $3 \leq n \leq 7$, we observed the following disadvantageous outcomes: (i) a large decomposition set of native gates, (ii) too many crossing connections among the physical qubits, i.e., many SWAP gates are added, and (iii) a higher depth (as the critical longest path along the native gates). From these disadvantageous outcomes, the quantum cost of the final transpiled quantum circuit is dramatically increased as *n* increases. For that, we designed the cost-effective *n*-bit Toffoli gate that has (i) a small set of IBM native single-qubit gates ($\sqrt{X}$ and RZ) and native double-qubit gates (CNOT), (ii) no crossing connections among the physical qubits, and (iii) a lower depth.

In this paper, this cost-effective *n*-bit Toffoli gate is termed the "layout-aware *n*-bit Toffoli gate". The construction of the layout-aware *n*-bit Toffoli gate mainly depends on the layout and the number of *n* physical qubits of an IBM QPU. This layout-aware *n*-bit Toffoli gate is entirely designed using the visual approach of the Bloch sphere, from the visual representations of the rotational quantum operations for the native ($\sqrt{X}$, RZ, and CNOT) gates. Moreover, we proposed a new formula for calculating the quantum cost for the final transpiled quantum circuit, and this formula is termed the "transpilation quantum cost (TQC)". The TQC calculates the total number of (i) the native single-qubit gates, (ii) the native double-qubit gates, (iii) the crossing connections, and (iv) the depth. After transpilation, using the aforementioned IBM QPUs for $3 \leq n \leq 7$, our proposed layout-aware *n*-bit Toffoli gate always has a lower TQC than that of the conventional *n*-bit Toffoli gate. In our experiments, we observed that the layout and the *n* physical qubits of an IBM QPU play two crucial roles in (i) minimizing the TQC and (ii) maintaining the structure of the layout-aware *n*-bit Toffoli gate, so that SWAP gates are not added. For that, we proposed the "optimal-*n*" and the "critical-*n*" parameters regarding these two crucial roles, when constructing the layout-aware *n*-bit Toffoli gate. For the previously stated IBM QPUs, we concluded that the two crucial roles are obtained when: (i) the optimal-*n* = 3 physical qubits for all layouts (linear, T-like, and I-like), and (ii) the optimal-*n* = ($3 \leq n \leq 4$) physical qubits for T-like and I-like layouts; otherwise, the two crucial roles are not obtained when the critical-*n* > 4 physical qubits for any layout (linear, T-like, or I-like). In conclusion, we proposed the layout-aware *n*-bit Toffoli gate to be utilized as a software package (transpilation library) for the IBM quantum system. Such that, when a developer transpiles a quantum circuit consisting of many conventional *n*-bit Toffoli gates, our proposed software package will transpile these many conventional gates to many layout-aware *n*-bit Toffoli gates, for a lower TQC with a chosen optimal-*n* parameter based on the layout and the number of *n* physical qubits of an IBM QPU.

### References

[1] A. A-Bayaty and M. Perkowski, "A concept of controlling Grover diffusion operator: A new approach to solve arbitrary Boolean-based problems," *Scientific Reports*, vol. 14, p. 23570, 2024.

[2] A. A-Bayaty and M. Perkowski, " BHT-QAOA: The generalization of quantum approximate optimization algorithm to solve arbitrary Boolean problems as Hamiltonians," *Entropy*, vol. 26, no. 10, p. 843, 2024.

[3] W. Li, Y. Ding, Y. Yang, R.S. Sherratt, J.H. Park, and J. Wang, "Parameterized algorithms of fundamental NP-hard problems: A survey," *Human-Centric Computing and Information Sciences*, vol. 10, no. 1, pp.1–24, 2020.

[4] L.K. Grover, "A fast quantum mechanical algorithm for database search," in *Proc. of 28th Annu. ACM Symp. on Theory of Comp.*, pp. 212–219, 1996.

[5] M. Perkowski, "Inverse problems, constraint satisfaction, reversible logic, invertible logic and Grover quantum oracles for practical problems," *Science of Computer Programming*, vol. 218, p. 02775, Jun. 2022.

[6] H. Simonis, "Sudoku as a constraint problem," in *CP Workshop on Modeling and Reformulating Constraint Satisfaction Problems*, vol. 12, pp. 13–27, 2005.

[7] H. Liu, F. Li, and Y. Fan, "Optimizing the quantum circuit for solving Boolean equations based on Grover search algorithm," *Electronics*, vol. 11, no. 15, p. 2467, 2022.

[8] M.A. Nielsen and I.L. Chuang, *Quantum Computation and Quantum Information*. 10th Anniversary ed., Cambridge University Press, ISBN 978-1107002173, 2010.

[9] R. LaPierre, *Introduction to Quantum Computing*. 1st ed., Springer, ISBN 978-3030693176, 2021.

[10] A. Mi, S. Deng, and J. Szefer, "Securing Reset Operations in NISQ Quantum Computers," in *Proc. of the 2022 ACM SIGSAC Conf. on Computer and Communications Security*, pp. 2279–2293, Nov. 2022.

[11] P. Murali, J.M. Baker, A. Javadi-Abhari, F.T. Chong, and M. Martonosi, "Noise-adaptive compiler mappings for noisy intermediate-scale quantum computers," in *Proc. of the 24th Int. Conf. on Architectural Support for Programming Languages and Operating Systems*, pp. 1015–1029, Apr. 2019.

[12] D. Koch, B. Martin, S. Patel, L. Wessing, and P.M. Alsing, "Demonstrating NISQ era challenges in algorithm design on IBM's 20 qubit quantum computer," *AIP Advances*, vol. 10, no. 9, 2020.

[13] P. Murali, N.M. Linke, M. Martonosi, A.J. Abhari, N.H. Nguyen, and C.H. Alderete, "Full-stack, real-system quantum computer studies: Architectural comparisons and design insights," in *Proc. of the 46th Int. Symp. on Computer Architecture*, pp. 527–540, Jun. 2019.

[14] A. Barenco, C.H. Bennett, R. Cleve, D.P DiVincenzo, N. Margolus, P. Shor, T. Sleator, J.A. Smolin, and H. Weinfurter, "Elementary gates for quantum computation," *Physical Review A*, vol. 52, no. 5, p. 3457, 1995.

[15] K.P. Gnatenko, H.P. Laba, and V.M. Tkachuk, "Energy levels estimation on a quantum computer by evolution of a physical quantity," *Physics Letters A*, vol. 424, p. 127843, 2022.

[16] T. Proctor, S. Seritan, K. Rudinger, E. Nielsen, R. Blume-Kohout, and K. Young, "Scalable randomized benchmarking of quantum computers using mirror circuits," *Physical Review Letters*, vol. 129, no. 15, p. 150502, 2022.

[17] IBM Quantum. "Transpiler." IBM.com. https://qiskit.org/documentation/apidoc/transpiler.html (accessed Jun. 15, 2023).

[18] V.V. Shende and I.L. Markov, "On the CNOT-cost of TOFFOLI gates," *arXiv preprint*, arXiv:0803.2316, 2008.

[19] B. Schmitt and G. De Micheli, "tweedledum: a compiler companion for quantum computing," in *Design, Automation & Test in Europe Conf. & Exhibition* (*DATE*), pp. 7–12, Mar. 2022. IEEE.

[20] B. Tan and J. Cong, "Optimal layout synthesis for quantum computing," in *Proc. of the 39th Int. Conf. on Computer-Aided Design*, pp. 1–9, Nov. 2020.

[21] M. Lukac, S. Nursultan, G Krylov, and O. Keszöcze, "Geometric refactoring of quantum and reversible circuits: Quantum layout," in *2020 23rd Euromicro Conf. on Digital System Design* (*DSD*), pp. 428–435, Aug. 2020. IEEE.

[22] Y. Takahashi and S. Tani, S., "Collapse of the hierarchy of constant-depth exact quantum circuits," *Computational Complexity*, vol. 25, pp. 849–881, 2016.

[23] A. Walker and D. Wood, "Locally balanced binary trees" *The Computer Journal*, vol. 19, no. 4, pp. 322–325, 1976.

[24] D. Maslov and G.W. Dueck, "Improved quantum cost for n-bit Toffoli gates," *Electronics Letters*, vol. 39, no. 25, pp. 1790–1791, 2003.

[25] J. Wang, G. Guo, and Z. Shan, "Sok: Benchmarking the performance of a quantum computer," *Entropy*, vol. 24, no. 10, p. 1467, 2022.

[26] B. Yang, R. Raymond, H. Imai, H. Chang, and H. Hiraishi, "Testing scalable Bell inequalities for quantum graph states on IBM quantum devices," *IEEE Journal on Emerging and Selected Topics in Circuits and Systems*, vol. 12, no. 3, pp. 638–647, 2022.

[27] M. Amico, H. Zhang, P. Jurcevic, L.S. Bishop, P. Nation, A. Wack, and D.C. McKay, "Defining standard strategies for quantum benchmarks," *arXiv preprint*, arXiv:2303.02108, 2023.

[28] IBM Quantum. "IBM Quantum access plans." IBM.com. https://www.ibm.com/quantum/access-plans (accessed Jun. 15, 2023).

[29] IBM Quantum. "IBM Quantum Systems." IBM.com. https://www.ibm.com/quantum/systems (accessed Jun. 15, 2023).

[30] IBM Quantum. "Qiskit Runtime." IBM.com. https://www.ibm.com/quantum/qiskit-runtime (accessed Aug. 26, 2023).

[31] P. Rao, K. Yu, H. Lim, D. Jin, and D. Choi, "Quantum amplitude estimation algorithms on IBM quantum devices," in *Quantum Communications and Quantum Imaging XVIII*, vol. 11507, pp. 49–60, Aug. 2020. SPIE.

[32] IBM Quantum. "System configuration." IBM.com. https://quantum-computing.ibm.com/lab/docs/iql/manage/systems/configuration (accessed Jun. 12, 2023).